\documentclass[12pt]{article}
\usepackage{graphicx}
\usepackage{epsfig}

\begin{document}

\title{Form factors in relativistic quantum mechanics:\\Is there a favored 
approach? Why? } 
\author{ 
B.  Desplanques\thanks{{\it E-mail address:}  desplanq@lpsc.in2p3.fr} \\ 
LPSC, Universit\'e Joseph Fourier Grenoble 1, CNRS/IN2P3, 
INPG,\\ 
  F-38026 Grenoble Cedex, France \\  }

\sloppy

\maketitle

\begin{abstract}
\small{Form factors of a simple system have been calculated in various forms of 
relativistic quantum mechanics, using a single-particle current. 
Their comparison has shown large discrepancies. 
The comparison is extended here to instant- and front-form calculations 
in unusual momentum configurations as well 
as to a point-form approach inspired from the Dirac's one (based on a 
hyperboloid surface). It is found that these new results depend on the 
momentum transfer, $Q$, through its ratio to the total mass, $Q/M$, 
(closely related to the Breit-frame velocity of the system). 
They evidence features similar for a part to those shown by 
an earlier ``point-form'' implementation (based on hyperplanes 
perpendicular to the velocity of the initial and final states). 
It thus appears that the standard instant- and front-form 
calculations, which generally do well compared either to experiment or to 
predictions of a theoretical model, rather represent exceptional cases. 
An argument explaining the success of these last approaches 
is presented and discussed. It is based on transformations 
of currents under Poincar\'e space-time translations, going beyond
the energy-momentum conservation property which results 
from the Lagrangian invariance under them. Depending on the approach, 
analytic or approximate numerical methods are proposed 
to correct form factors for missing constraints then expected.
}
\end{abstract} 


\newpage

\section{Introduction}
Relativistic quantum mechanics (RQM) can be approached by different forms,  
originally classified by Dirac~\cite{Dirac:1949cp}. 
When calculating properties of hadrons like form factors, using 
a single-particle current, front and instant ones are often 
considered as more appropriate. In the case of the pion charge form factor 
for instance, these approaches 
\cite{Isgur:1984jm,Chung:1988mu,Cardarelli:1995dc,Choi:1997iq,deMelo:1997cb}
provide results relatively close to experiment
\cite{Bebek:1978pe,Amendolia:1986wj,Volmer:2000ek}.  
A similar statement holds in the case of theoretical models involving 
spinless constituents~\cite{Amghar:2002jx}. 
It is not always stressed however that these calculations correspond 
to a particular momentum configuration, $q^+=0$ in the former case 
and Breit frame in the latter. 

Recently, an implementation of the point-form approach \cite{Klink:1998} 
has been employed for the calculation of form factors of hadronic systems 
such as the pion \cite{Allen:1998hb,Amghar:2003tx,He:2004ba}, the deuteron 
\cite{Allen:2000ge} and the nucleon~\cite{Wagenbrunn:2000es}. In the pion case, 
which has been revisited~\cite{Amghar:2003tx,He:2004ba}, the approach produces 
a huge discrepancy with experiment. A similar statement could be made 
in the case of theoretical models involving a two-body system with equal-mass 
constituents~\cite{Amghar:2002jx,Desplanques:2001zw}, 
especially when the total mass of the system, $M$, becomes small 
in comparison with the sum of the constituent masses, $2\,m$. 
The discrepancy can be traced back to the dependence of form factors 
on the momentum transfer $Q$ through the velocity of the system 
in the Breit frame for instance, which involves the ratio $Q/(2M)$. 
This ratio, which is the only parameter entering the boost transformation 
in the point-form approach, has striking consequences. The charge radius 
varies like the inverse of the mass of the system with the result that 
the larger the binding, the larger the radius, which is somewhat 
counter-intuitive. Moreover, at high momentum transfer, the power-law 
behavior of the form factor evidences a suppression by as many powers 
of $M/(2\,m)$ as there are powers of $Q$. The appearance in the point-form 
approach of the total mass of the system, $M$, and its essential role, 
especially when it goes to zero, were emphasized in various works 
\cite{Desplanques:2001ze,Desplanques:2001zw,Amghar:2003tx,Coester:2004qu}. 
At the same time, it was noticed that this mass was playing no role 
in the boost transformation required for calculating form factors 
in the standard instant- and front-form approaches (as far as 
a single-particle current is concerned).

When the first ``point-form'' result evidencing a sizeable discrepancy 
appeared \cite{Desplanques:2001zw}, it could be thought that this 
was a peculiarity of the approach \cite{Klink:1998}, suggesting 
a specific problem. It was thus found that form factors so obtained 
systematically  evidence wrong power-law behavior at high $Q^2$. 
This problem could be solved by introducing the simplest two-body 
currents \cite{Desplanques:2003nk}. These ones could not however remove 
other important drawbacks evidenced by the approach.

Interestingly, results showing similarities with the above ``point-form'' 
results have recently appeared in other approaches, as a by-product 
of studies aiming to look at the frame dependence of form factors and 
the role of two-body currents. Thus, an instant-form calculation of the 
form factor for a strongly bound system showed a fast fall-off when 
going away from the Breit frame \cite{Amghar:2002jx}. In a front-form 
approach, Simula examined results for the form factor of a pseudo-scalar 
meson with the pion mass \cite{Simula:2002vm}. In the case $q^+\neq0$, 
a large drop-off of the single-particle contribution was observed 
while the relation of the effect to the dependence of the form 
factor on the ratio $Q/(2M)$ was emphasized. A similar effect was 
also found in field-theory motivated approaches, with $q^+\neq0$ 
\cite{Bakker:2000pk,deMelo:2002yq}. No relation to a 
dependence of the corresponding contribution on the ratio $Q/(2M)$ 
was made in these last cases but, in view of the results, there 
is not much doubt on the origin of the effect. Thus, the striking behavior 
of the single-particle current contribution to form factors 
of strongly bound systems in the earlier ``point-form'' approach is far 
to be an isolated fact. 

The above observation has been obtained in different schemes however. 
Its general character needs to be confirmed and specified by 
dedicated studies involving the same inputs as much as possible. 
Though we believe that there is some relationship between effects 
in field-theory and RQM approaches, the last ones have their own rules 
and a full correspondence is not guaranteed. It is therefore appropriate 
to extend earlier comparisons of form factors obtained in different forms 
of relativistic quantum mechanics \cite{Amghar:2002jx}. Given that 
form factors calculated in the front and instant forms are not 
Lorentz invariant, one can thus look at them for non-standard momentum
configurations. 
The comparison can provide some information on violations of Lorentz 
invariance but this supposes that no other symmetry is significantly 
violated at the same time. One can also consider a point-form approach 
more in the spirit of the Dirac's one, based on a hyperboloid surface 
\cite{Desplanques:2004rd}. 
Apart from the fact that such an approach has never been used, it can provide 
information on the specific character  of results obtained in the earlier 
``point form'' implementation. As noticed by Sokolov  \cite{Sokolov:1985jv}, 
this one implies a hyperplane perpendicular to the velocity of the system and 
therefore differs from the Dirac's point form. Though these features are not so
clearly expressed, they also stem from an earlier work by Bakamjian
\cite{Bakamjian:1961} where it is
shown that ``an instant form of relativistic quantum mechanics can be
constructed which displays the symmetry properties inherently present in the
point form''.  The two approaches (hyperplane and hyperboloid based) have 
in common that only the generators $P^{\mu}$ of the Poincar\'e algebra contain 
the interaction. To distinguish them, we use quotation marks when referring to 
the first one.  

Ultimately, one would like to get a sufficiently large insight 
on form factors in different forms and  different momentum configurations 
so that it can 
provide a clue as to why calculations based on a single-particle current do 
relatively well in some approaches while they cannot in other ones. 
For these last cases, a major role is played by two-body currents which, 
in principle, should ensure that form factors be independent of the form 
and frame under consideration. Evidently, a comparison to the 
predictions of an underlying field-theory model, as done in Refs. 
\cite{Desplanques:2001zw,Amghar:2002jx}, can tell about  
the efficiency of an approach. However, apart from the fact that 
this is not always possible, we believe that a sensible argument, 
if any, should be found within the formalism that is employed. 
In RQM approaches, the description of the initial and final states entering the
calculation of form factors fulfills Poincar\'e covariance properties by
construction of the corres\-pon\-ding algebra. This involves (homogeneous) Lorentz
transformations (boosts and rotations) generated by the algebra operators,
$M^{\mu\,\nu}$, and space-time translations generated by the 4-momentum
operators, $P^{\mu}$. However, at the interaction vertex of the external probe
with the constituents of the system under consideration, 
Poincar\'e covariance is generally violated. 
Depending on the approach, only part of the expected properties 
is fulfilled. Those related to Lorentz transformations, which can be easily
checked by mo\-ving or rotating the system, are currently emphasized. 
On the contrary, due to the absence of a similar check, 
the properties related to space-time translations, 
beyond the global 4-momentum conservation that results from 
the Lagrangian invariance under these transformations, 
are essentially unexplored.
It is our intent in this paper to show that these properties 
can play an important role in discriminating various results. 

The study is done for the ground state of a two-body system 
with equal-mass cons\-tituents, in a scalar-particle model. This offers 
the advantage of minimizing specific difficulties pertinent to the 
description of a more realistic system like a hadronic one, due to the 
non-zero spin of the constituents or to a complicated dynamics which 
one would generally like to learn about.  Beside results of approaches 
currently considered in the literature, we consider new ones which involve, 
on the one hand, extensions of instant- and front-form approaches 
to a ``parallel'' momentum configuration most often ignored 
and, on the other hand, a point-form approach inspired from Dirac's one 
\cite{Desplanques:2004rd}. In the last case, the calculation
of form factors supposes some elaboration. 
They represent a  straightforward generalization of those obtained 
in the front form with an arbitrary orientation of the front. 
Demonstrating their Lorentz invariance is more tedious however. 
Together with earlier results, the new ones turn out to be important 
in revealing both the respective merit of different approaches 
and the role of properties related to space-time translations.

The plan of the paper is as follows. After reminding some generalities 
relative to the ingredients entering the calculation of form factors 
in relativistic quantum mechanics, we successively consider 
in the second section expressions of form factors in front 
and instant forms for unusual momentum configurations, 
and in a point form  closer to Dirac's one. The section is ended 
by the consideration of form factors in a field-theory model, 
which in some sense play here the role of an ``experiment''. 
Results are presented in the third section. They involve two 
form factors (Lorentz vector and scalar) while attention 
is given to both the low- and high-$Q^2$ behavior, 
in relation respectively with the radius and power law expectations. 
A discussion of the results in the light of transformation properties 
of currents under Poincar\'e space-time translations
is given in the fourth section. The conclusion follows in the fifth section. 
An appendix contains many details relevant to the derivation 
of form factors in the ``parallel'' momentum configuration 
(with $\bar{P}\rightarrow \infty$) and in the Dirac's 
inspired point form. A part is devoted to corrections that allow one to get  
a Lorentz-invariant scalar form factor at $Q^2=0$ in the instant form.

\section{Expression of form factors in different approaches}
\begin{figure}[htb]
\begin{center}
\mbox{ \psfig{ file=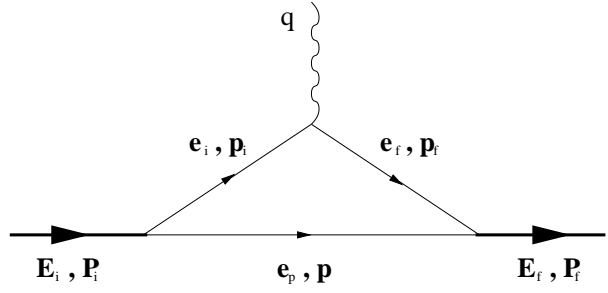, width=8cm}}
\end{center}
\caption{Photon absorption on a two-body system: kinematics relative to a 
RQM approach (particles on-mass shell:
$e_p=\sqrt{m^2+p^2}$). Our convention assumes $P_f^{\mu}=P_i^{\mu}+q^{\mu}$ 
\label{fig1} } 
\end{figure}  
For the ground state of the system considered here, made of scalar particles, 
there are two form factors, $F_1(Q^2)$ and $F_0(Q^2)$, corresponding 
to a vector and a scalar probe respectively. Their general definition may 
be found in Ref. \cite{Amghar:2002jx} while a schematic representation of the 
contribution in the single-particle approximation is given in Fig. \ref{fig1}. 
Considering both of them can provide a better insight on their properties. 
Their determination in relativistic quantum mechanics implies two ingredients: 
the relation of the constituent momenta to the total momentum and the solution 
of a mass operator. These ingredients and the corresponding form factors are 
successively considered in what follows.

\subsection{Wave functions}
In all cases we are considering, the relation of the constituent momenta, 
$\vec{p_1}$ and $\vec{p_2}$, to the total momentum, $\vec{P}$, can be cast 
into a unique form:
\begin{equation}
\vec{p_1}+\vec{p_2}-\vec{P}=\frac{\vec{\xi}}{\xi^0}\,(e_1+e_2-E_P),
\label{boost0}
\end{equation}
where the 4-vector, $\xi^{\mu}$, characterizes each approach. Following 
the work underlying the Bakamjian-Thomas construction of the Poincar\'e 
algebra in the instant form \cite{Bakamjian:1953kh}, it is appropriate to 
introduce a Lorentz-type transformation that allows one to express the 
constituent momenta in terms of an internal variable, $\vec{k}$, 
which enters the mass operator, and the total momentum, $\vec{P}$. 
This transformation, which preserves the on-mass shell character of 
constituents while fulfilling Eq. (\ref{boost0}), is given by:
\begin{equation}
\vec{p}_{1,2} = \pm\,\vec{k} \pm \vec{w}\;
\frac{\vec{w} \! \cdot \!\vec{k} }{w^0 +1}
 + \vec{w}  \;e_k \, ,\hspace{1cm}
e_{1,2} = w^0\,e_k\pm \vec{w}\! \cdot \!
 \vec{k}   \,,
\label{boost1}
\end{equation}
where the $\vec{k}$ vector is defined up to a rotation\footnote{This
indetermination has no effect on the calculation of form factors. It could
however affect, for instance, the comparison of integral expressions aiming 
to the calculation of the same quantity in different frames when given 
by an integral over the $\vec{k}$ variable.} while the components of the 4-vector $w^{\mu}$, $\vec{w}$ and 
$w^0=\sqrt{1+(\vec{w})^2}$, are given by: 
\begin{equation}
 w^{\mu} =\frac{ P^{\mu}}{ 2\,e_k }
+ \frac{ \xi^{\mu} }{ 2\,e_k } \; \frac{4\,e_k^2-M^2}{
 \sqrt{ (\xi \cdot P)^2 + (4\,e_k^2-M^2)\;\xi^2 } +  \xi \cdot P} \,.
\label{boost2} 
\end{equation}
The above details, pertinent to the Bakamjian-Thomas construction of the
Poincar\'e algebra, can often be skipped for practical purposes. With this
respect, relations of interest, which in particular are independent of the
orientation of the $\vec{k}$-vector, are the following ones:
\begin{equation}
(p_1+p_2)^2=4\,e_k^2\, ,
\label{ek2} 
\end{equation}
\begin{equation}
(p_1+p_2)^{\mu}=P^{\mu}
+  \xi^{\mu}  \; \frac{4\,e_k^2-M^2}{
 \sqrt{ (\xi \cdot P)^2 + (4\,e_k^2-M^2)\;\xi^2 } +  \xi \cdot P} \, .
\label{pisupf} 
\end{equation}

It is noticed that the 4-vector, $\xi^{\mu}$, appearing in the above 
expressions, is always associated with a factor $(4\,e_k^2-M^2)$, which 
is nothing but an interaction term. The appearance of this one is a 
consequence of relying on a unique hypersurface to describe the physics, 
independent of the system under consideration.
It is also seen that the expression is independent of the scale 
of the 4-vector $\xi^{\mu}$. Thus, up to an irrelevant scale,  the 4-vector 
$\xi^{\mu}$, which reflects the symmetry properties of the hypersurface 
underlying each approach, is given as follows:\vspace{3mm}\\
$\bullet$ {\it instant form}
\begin{equation}
\xi^0=1, \;\;\;  \;\vec{\xi}=0\,,
\label{boost3} 
\end{equation}
$\bullet$ {\it front form}
\begin{equation}
\xi^0=1, \;\;\;  \;\vec{\xi}=\hat{n}\,,
\label{boost4} 
\end{equation}
\hspace{3mm} 
where $\hat{n}$ is a unit vector with a fixed direction ($\xi^2=0$),
\vspace{3mm}\\
$\bullet$ {\it Dirac's inspired point form} 
\cite{Desplanques:2004rd}
\begin{equation}
\xi^0=u^0=1, \;\;\;  \;\vec{\xi}=\hat{u}\,,
\label{boost5} 
\end{equation}
\hspace{3mm} 
where $\hat{u}$ is a unit vector that points to any direction ($\xi^2=0$). 

The above equations, (\ref{boost1}) and (\ref{boost2}), can be generalized to
an arbitrary hyperplane with orientation $\xi^{\mu}=\lambda^{\mu}$ and
$\lambda^2=1$. They, in particular,  allow one to recover 
the boost transformation introduced in an earlier ``point-form'' approach 
\cite{Klink:1998,Bakamjian:1961}  by taking $\xi^{\mu} \propto P^{\mu}$. 
The corresponding 
expression of the $w^{\mu}$  four vector, $ w^{\mu}=P^{\mu}/M$, can 
be obtained from the other cases by neglecting interaction effects 
($2\,e_k \rightarrow M$). Missing the consistency requirement that 
underlies them, the calculation of form factors in this approach 
ne\-ces\-sarily implies two hyperplanes determined by the different momenta 
of the initial and final states.  

\begin{figure}[htb]
\begin{center}
\mbox{ \psfig{ file=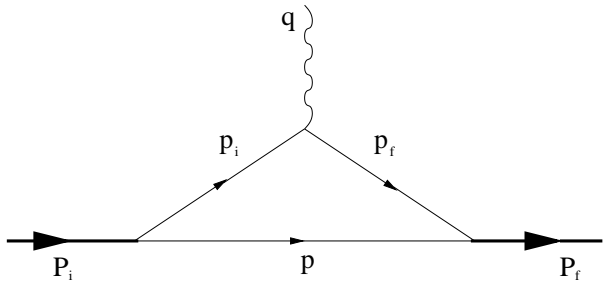, width=8cm}}
\end{center}
\caption{Photon absorption on a two-body system: Feynman diagram with 
corresponding kinematics. 
\label{fig2} }
\end{figure}  
The second main ingredient entering the calculation of form factors 
concerns a mass operator and its solution,  which could be written as:
\begin{equation}
(M^2-4\,e_k^2)\;\phi_0(\vec{k})=
\int \frac{d\vec{k}\,'}{(2\pi)^3} \; 
\frac{1}{ \sqrt{e_{k}}  }\; V_{int}(\vec{k},\vec{k}\,') \;
\frac{1}{ \sqrt{e_{k'}} } \; \phi_0(\vec{k}\,') \, .
\label{wf6}
\end{equation} 
A particular case corresponds to the Feynman diagram shown in Fig. \ref{fig2}. 
The strong interaction vertices appearing in this diagram can be considered as 
resulting from the exchange of an infinite-mass boson. Due to this property, 
little uncertainty on the determination of the mass operator and its solution 
is expected. Actually, an exact solution can be found \cite{Amghar:2002jx}. 
It  only involves the masses of the constituents, $m$, and of the system, $M$. 
It is given by:
\begin{equation}
\phi_0(\vec{k})=\phi_0(k)
 \propto \frac{1}{ \sqrt{e_{k}} }\; \frac{1}{ 4\,e_k^2-M^2}.
\label{wf7}
\end{equation} 

Another extreme case of interest corresponds to the exchange of a zero-mass 
boson (Wick-Cutkosky model). Due to a hidden symmetry, the solution of the 
Bethe-Salpeter equation  can be obtained relatively easily. As shown in Refs. 
\cite{Amghar:2002jx,Amghar:2000pc}, a reasonable mass operator can be  
determined. 

For practical purposes, we take for the constituent masses and the total mass 
of the system under consideration values appropriate to the pion, $m=0.3$ GeV 
and $M=0.14$ GeV. For the zero-mass boson exchange case, we rely on a mass 
operator of the form given in Eq. (56) of Ref. \cite{Amghar:2002jx} with 
a coupling constant determined by fitting the total mass. 
Its high $k$ behavior is essential in getting the appropriate 
power-law behavior of form factors. The following normalization of the 
ground-state wave function of interest here is assumed:
\begin{equation}
\int \frac{ d\vec{k}}{(2\pi)^3 } \; \phi^2_0(k)=1.
\label{wf8}
\end{equation} 
%

\subsection{Form factors in hyperplane-based approaches}
In order to calculate form factors, the current to be used has to be specified. 
This was done in Ref. \cite{Amghar:2002jx} where their expressions in various 
approaches involving hyperplanes (instant, front  and earlier point form) have 
been given. Written in terms of different variables, it was not always obvious 
how similar they were looking like. 

In the case of the charge form factor, $F_1(Q^2)$, where the similarity 
is the most striking, it is found that the different form factors 
can be cast into a common form given by:
\begin{eqnarray}  
F_1(Q^2)=   
\frac{1}{(2\pi)^3 }  
\int \frac{ d\vec{p}}{2\,e_p} 
\;\Big( (p_f\!+\!p)^2 \, (p_i\!+\!p)^2 \Big)^{1/4} \;
\phi_0 \Big((\frac{p_f\!-\!p}{2})^2\Big)  \;    
\phi_0 \Big( (\frac{p_i\!-\!p}{2})^2 \Big)  
  \nonumber  \\  \times 
\;\;\frac{\xi_f \cdot (p_f+p)\;\;\xi_i \cdot (p_i+p)}{
(2\,\xi_f \cdot p_f) \;(2\,\xi_i \cdot p_i) }\;\;
\frac{2\,(p_f+p_i) \cdot (\xi_f+\xi_i )}{
(p_f+p_i+2\,p) \cdot  (\xi_f+\xi_i )}\,.\hspace{2cm}
  \label{ff1}
\end{eqnarray}
The relation of the 4-momenta $p_i^{\mu}$ or $p_f^{\mu}$ to the 4-momentum 
of the spectator particle $p^{\mu}$, the total momenta $P_i^{\mu}$ or 
$P_f^{\mu}$ and the 4-vectors $\xi_i^{\mu}$ or $\xi_f^{\mu}$ is given 
by Eq. (\ref{boost0}) and subsequent ones. The argument of the wave function, 
$((p_1\!-\!p_2)/2)^2$ is related to the internal $\vec{k}$ variable 
by the relation $((p_1\!-\!p_2)/2)^2=-\vec{k}\,^2$, so that 
$\phi_0 \big(((p_1\!-\!p_2)/2)^2\big)$ in Eq. (\ref{ff1}) stands for 
$\phi_0 \big(k=((-(p_1\!-\!p_2)/2)^2)^{1/2}\big)$.
The 4-vectors $\xi_i^{\mu}$ and $\xi_f^{\mu}$ are identical for the instant 
and front forms and are given by Eqs. (\ref{boost3}) and (\ref{boost4})  
respectively. In the front-form case, with the condition $q^+=0$, standard 
expressions in terms of the $x$ and $k_{\perp}$ variables can be recovered 
by making a change of variables (the demonstration is similar to that one 
given in Appendix \ref{app:a} for a different momentum configuration). 
For the earlier ``point form'', the above 4-vectors are proportional 
to the velocity of the initial and final states and are taken as 
$\xi_i^{\mu}=P_i^{\mu}/M$, $\xi_f^{\mu}=P_f^{\mu}/M$. As expected, 
Eq. (\ref{ff1}) is invariant under a change of the scale 
of the  4-vectors, $\xi^{\mu}$. It can also be shown that the following 
equality holds:
\begin{equation}
F_1(Q^2=0)=\int \frac{ d\vec{k}}{(2\pi)^3 } \; \phi^2_0(k)=1,
\label{ff2}
\end{equation} 
independently of the velocity of the system. The specific form of the last 
factor at the second line in Eq. (\ref{ff1}) is especially relevant 
to obtain this result in the instant form.

In the case of the scalar form factor, $F_0(Q^2)$, the expression of the
different form factors reads:
\begin{eqnarray}  
F_0(Q^2)=   
\frac{1}{(2\pi)^3 }  
\int \frac{ d\vec{p}}{2\,e_p} 
\;\Big( (p_f\!+\!p)^2 \, (p_i\!+\!p)^2 \Big)^{1/4} \;
\phi_0 \Big((\frac{p_f\!-\!p}{2})^2\Big)  \;    
\phi_0 \Big( (\frac{p_i\!-\!p}{2})^2 \Big)  
  \nonumber  \\  \times 
\;\;\frac{\xi_f \cdot (p_f+p)\;\;\xi_i \cdot (p_i+p)}{
(2\,\xi_f \cdot p_f) \;(2\,\xi_i \cdot p_i) }\;\;cf_0\,,\hspace{5cm}
  \label{ff3}
\end{eqnarray}
where the definitions relative to momenta are the same as for $F_1(Q^2)$. 
The coefficient, $cf_0$, is introduced to make the value of the scalar form
factor at $Q^2=0$, $F_0(0)$, independent of the velocity of the system, as
expected from a minimal Lorentz invariance requirement.  With this respect, 
only the instant-form approach raises a problem. The method allowing 
to determine $cf_0$ is described in the Appendix \ref{app:b}. Its expression
reads:
\begin{equation}
cf_0=1+\frac{g(k_i)+g(k_f)}{2}\;\;
\frac{e_p \,\Big((e_f+e_p)(e_i+e_p)-E_f\;E_i\Big)}{(e_f+e_p)(e_i+e_p)(e_f+e_i)},
\label{ff4}
\end{equation}
where the function $g(k)$ can be obtained from a quadrature, 
see Eq. (\ref{appb4}). In the case of a scalar-particle model 
together with a zero-range interaction, it is found that $g(k)=1$. 
The above factor then allows one to reproduce the ``exact'' scalar 
form factor, $F_0(Q^2)$, at all $Q^2$. In the other extreme corresponding 
to a zero-mass exchange interaction (Wick-Cutkosky model),  $g(k)$ takes 
a value close to $1/3$. In all the other approaches, we assume $cf_0=1$. 
Such a result is actually obtained from an extension of Eq. (\ref{ff4}) 
to an arbitrary hyperplane with orientation given by a 4-vector $\xi^{\mu}$:
\begin{equation}
cf_0=1+\frac{g(k_i)+g(k_f)}{2}\;\;
\frac{\xi \!\cdot\! p \;\Big(\xi \!\cdot\! (p_f+p)\;\;\xi \!\cdot\! (p_i+p)-
   \xi \!\cdot\! P_f\;\xi \!\cdot\! P_i\Big)
}{\xi \!\cdot\! (p_f+p)\;\;\xi \!\cdot\! (p_i+p) \;\;\xi \!\cdot\! (p_f+p_i)}.
\label{ff4bis}
\end{equation}
For the front form (using Eq. (\ref{boost0})) or the instant form with the
``parallel'' momentum configuration 
and $\bar{P} \rightarrow \infty$ (see Eq. (\ref{appa2})), 
it is found that the coefficient of the $g(k)$ factor in the above equation 
vanishes.

It is noticed that the last factor in Eq. (\ref{ff1}) characterizes 
the charge form factor. The numerator contains the factor $(p_i+p_f)^{\mu}$ 
which is part of the photon coupling to scalar particles while 
the denominator corresponds to the factor $(P_i+P_f)^{\mu}$ that has 
to be factored out from the matrix element of the current. 
Its particular form ensures that the ratio $F_1(0)/F_0(0)$ 
predicted by simple theoretical models is approximately or exactly 
recovered \cite{Amghar:2002jx}. The last factor  in Eq. (\ref{ff1}) 
is evidently absent for the scalar form factors where it is replaced 
by the factor $cf_0$ when necessary (instant form), ensuring that 
the form factor at $Q^2=0$ be Lorentz invariant as already explained. 
This amounts to account for some two-body currents. We would finally like 
to remark that the above factor in Eq. (\ref{ff1}) is responsible for getting 
the asymptotic ratio, $F_1(Q^2)/F_0(Q^2)\;(Q^2 \rightarrow \infty)=2$  
(taking into account the definitions adopted for the form factors
\cite{Amghar:2002jx}). 
This agrees with expectations from the underlying field-theory model.

\subsection{Limit for a parallel-momentum configuration (and large momenta)}
It is known that the expression of a form factor calculated in the instant form 
with the momentum configuration $\vec{q} \perp (\vec{P}_i+\vec{P}_f)\; (E_i=E_f)$ and 
$|\vec{P}_i+\vec{P}_f| \rightarrow \infty$ is close or even identical 
to that one obtained in the standard front-form approach ($q^+=0$, 
also denoted ``perpendicular'' in the following). The choice 
of the currents ensures the identity in the present case. Another limit 
of interest corresponds to take the parallel configuration, 
$\vec{q} \parallel (\vec{P}_i+\vec{P}_f)\; (E_i \neq E_f)$ and 
$|\vec{P}_i+\vec{P}_f| \rightarrow \infty$ (denoted ``parallel'' 
in the following). Apart from a few recent works, 
this limit is rarely considered. Form factors so obtained however evidence 
a feature that, in our opinion, casts a completely new insight on earlier 
``point-form'' results. We give here the expression of the form factors 
in this limit while some details about the derivation are given 
in Appendix \ref{app:a}. This is conveniently done using the variables 
employed in the standard front-form approach, $x$ and $k_{\perp}$. 
The noticeable point is that the dependence on the momentum transfer, $Q$,
appears through the quantity $v= \sqrt{Q^2/(Q^2+4\,M^2)}$, which is 
nothing but the velocity of the system in the Breit frame and, 
most important, is the same as in point-form approaches. 
The most symmetrical expressions read:
\begin{eqnarray}
F_1(Q^2)&=& \frac{1}{(2\,\pi)^3} \int_0^{1-v} dx \;\frac{ (1-x)}{2\,x}\; 
\frac{1-v^2}{(1-x)^2-v^2} \;\int d^2k_{\perp} \;
\tilde{\phi}(k_i^2)\;\tilde{\phi}(k_f^2),
\nonumber \\
F_0(Q^2)&=& \frac{1}{(2\,\pi)^3} \int_0^{1-v} dx \;\frac{ 1}{4\,x} \; 
\frac{1-v^2}{(1-x)^2-v^2} \;\int d^2k_{\perp} 
\;\tilde{\phi}(k_i^2)\;\tilde{\phi}(k_f^2),
\nonumber \\
{\rm with} \hspace{6mm}
\tilde{\phi}(k)&=&\sqrt{e_k}\;\phi_0(k),
\nonumber \\  
k_i^2&=&k^2_{\perp} +(m^2+k^2_{\perp}) \; \frac{(1-2\,x-v)^2}{4\,x\,(1-x-v)},
\nonumber \\  
k_f^2&=&k^2_{\perp} +(m^2+k^2_{\perp} )\; \frac{(1-2\,x+v)^2}{4\,x\,(1-x+v)}\,.
\label{ff5}
\end{eqnarray}
Interestingly, these expressions are identical to those obtained in the front 
form with the momentum configuration $\vec{q} \parallel \vec{n}$ 
(denoted ``parallel'' in the following), which  can be obtained 
from the original ones,  Eqs. (\ref{ff1}, \ref{ff3}), by performing 
a change of variable (see second part of Appendix \ref{app:a}). 
It is reminded that a similar identity generally holds for 
the ``perpendicular'' momentum configuration. 

\subsection{Form factor in Dirac's point form case}
The need for developing a point-form approach more in the spirit 
of the Dirac's one \cite{Desplanques:2004rd} is due for a part to the drawbacks 
evidenced by an earlier implementation to reproduce form factors calculated 
in a very simple theoretical model \cite{Desplanques:2001zw}. As mentioned 
elsewhere \cite{Desplanques:2001ze}, this last approach implies hyperplanes
perpendicular to the velocity of the initial and final states while 
the Dirac's one is based on a hyperboloid surface, which is 
at the same time unique and independent of the system under consideration. 
The question therefore arises of whether the new approach can improve 
the calculation of form factors. In Ref. \cite{Desplanques:2004rd}, 
it is shown that the correct power-law behavior of form factors could be 
recovered in a simple case. We here consider the calculation of these form 
factors on a more general ground, which is done for the first time. 
We give and explain the expression of form factors in this subsection 
while details are given in Appendix \ref{app:c}. 

Consistently with the absence of a direction on a hyperboloid, it is expected 
that the sum of the constituent momenta points isotropically to any 
direction in the c.m. case. This direction, which is to some extent 
a new degree of freedom and turns out to be conserved, is here 
represented by a unit vector $\hat{u}$. It is therefore expected that the 
expression of form factors involves an integration over this orientation. 
This represents the main new feature evidenced by the form factors considered 
here. The integration over $\hat{u}$ is not arbitrary however. 
It includes some weight which ensures Lorentz invariance and is obtained from 
considering the expression of the norm for instance. A minimal expression 
thus takes the form:
\begin{equation}
F(Q^2)=\int \frac{d\hat{u}}{4\,\pi} \;\Big(\frac{M}{P \cdot u}\Big)^2 \cdots\, ,
\label{ff6}
\end{equation}
where $P^{\mu}$ is the 4-momentum of the initial or final state. How factors 
entering Eq. (\ref{ff6}) change under a Lorentz transformation while 
preserving the invariance of the full expression is described in 
Appendix \ref{app:c}. Two points in the demonstration are to be noticed. 
The property, $(u^0)^2- \vec{u}\,^2=0$, remains unchanged, allowing one to
define a new unit vector, $\hat{u}\,'=\vec{u}/u^0$. The change 
in the scale of the 4-vector, $u^{\mu}$, which occurs in the transformation, 
is compensated by a modification in the other factors so that we can 
choose $u'\,^0=1$, while the change in the 
orientation, $\hat{u} \rightarrow \hat{u}\,'$, can be absorbed into 
the integration over $\hat{u}$. As for the dots at the r.h.s. 
of Eq. (\ref{ff6}), they represent other  factors like wave functions 
and matrix elements of the current. The important point is that these 
factors be formally Lorentz invariant and invariant under a change 
of scale of the 4-vector, $u^{\mu}$. For this part, one can thus use 
Eqs. (\ref{ff1}, \ref{ff3}), which offers the advantage to introduce no bias in the 
comparison that will be made later on with other forms. 
Taking into account the symmetry between initial and final states, the 
expressions for the charge and scalar form factors thus read:
\begin{eqnarray}  
F_1(Q^2)&=&  \int \frac{d\hat{u}}{4\,\pi} \;\Big(\frac{M}{P_i \cdot u}\Big)^2
\frac{1}{(2\pi)^3 }  
\int \frac{ d\vec{p}}{2\,e_p} 
\;\phi_0 \Big((\frac{p_f\!-\!p}{2})^2\Big)  \;    
\phi_0 \Big( (\frac{p_i\!-\!p}{2})^2 \Big)  
  \nonumber  \\   &&\hspace*{-1cm} \times \;
\Big( (p_f\!+\!p)^2 \, (p_i\!+\!p)^2 \Big)^{1/4} \;
\;\;\frac{u \cdot (p_f\!+\!p)\;\;u \cdot (p_i\!+\!p)}{
(2\,u \cdot p_f) \;(2\,u \cdot p_i) }\;\;
\frac{2\,(p_f\!+\!p_i) \cdot u}{
(p_f\!+\!p_i\!+\!2\,p) \cdot u}\;, 
\nonumber \\
F_0(Q^2)&=&   \int \frac{d\hat{u}}{4\,\pi}\; \Big(\frac{M}{P_i \cdot u}\Big)^2 
\frac{1}{(2\pi)^3 }  
\int \frac{ d\vec{p}}{2\,e_p} 
\;\phi_0 \Big((\frac{p_f\!-\!p}{2})^2\Big)  \;    
\phi_0 \Big( (\frac{p_i\!-\!p}{2})^2 \Big)  
  \nonumber  \\  && \times \;
\Big( (p_f\!+\!p)^2 \, (p_i\!+\!p)^2 \Big)^{1/4} \;
\;\;\frac{u \cdot (p_f\!+\!p)\;\;u \cdot (p_i\!+\!p)}{
(2\,u \cdot p_f) \;(2\,u \cdot p_i) }\;.
\label{ff7}
\end{eqnarray}  
In these expressions, the relation of the struck-particle momenta, 
$p_i^{\mu},\;p_f^{\mu}$, are given in terms of the spectator one, $p^{\mu}$, 
by Eqs. (\ref{boost0}-\ref{boost2}) and (\ref{boost5}).

\subsection{``Experiment''}
Among theoretical models, two of them, already mentioned,  are of special 
interest. Corresponding to two opposite extreme cases, they involve scalar 
particles interacting by exchanging an infinite-mass boson (interpretation of 
strong interaction vertices in the triangle Feynman diagram, Fig. \ref{fig2}) 
and a zero-mass one (Wick-Cutkosky model). In both cases, the Bethe-Salpeter 
amplitude is given under a form which is analytical for the essential factors. 
The calculation of form factors, which implies a Wick rotation, can thus be 
performed without much difficulty. Expressions of charge and scalar 
form factors so obtained for the ground state have been given  in Eq. (48) of 
Ref. \cite{Amghar:2002jx} for the triangle Feynman diagram and the appendix of
Ref. \cite{Desplanques:2003nk} for the Wick-Cutkosky model. 
We here notice that these form factors behave asymptotically as $Q^{-2}$ 
and $Q^{-4}$ respectively, up to log terms (see Alabiso and Schierholz for 
more general predictions in relation with the behavior of the interaction at
high-momentum transfer \cite{Alabiso:1974sg}). In the former case, an analytic 
expression is available in the limit $ M \rightarrow 0$ (see
Ref. \cite{Amghar:2002jx}, Eq. (50)). In the latter one, only an approximate 
expression can be obtained in the same limit. The asymptotic behavior 
can nevertheless be determined exactly. As it offers some interest (beside
providing a useful numerical check), it is given here: 
\begin{eqnarray}  
F_1(Q^2)_{Q^2 \rightarrow \infty}&=&
 2\,F_0(Q^2) _{Q^2 \rightarrow \infty} 
 \nonumber \\
&=& 540\,(\frac{m^2}{Q^2})^2\; \Bigg(\frac{\sqrt{4\,m^2+Q^2}}{2\,Q}\;
{\rm log} \Bigg[\frac{ \sqrt{4\,m^2+Q^2}+Q }{ \sqrt{4\,m^2+Q^2}-Q } \Bigg] 
-2 
 \nonumber \\ && \hspace{2.5cm}
+ \frac{m^2}{Q^2}\; 
{\rm log}^2 \Bigg[\frac{ \sqrt{4\,m^2+Q^2}+Q }{ \sqrt{4\,m^2+Q^2}-Q } \Bigg]
\Bigg).
\label{asym0}
\end{eqnarray}  
Departures to the exact result, located at low $Q^2$, do not exceed those
observed at $Q^2=0$ (3/2 and 3/4 instead of 1 and 5/4 for $F_1(0)$ 
and $F_0(0)$ respectively). Departures due to the non-zero mass of $M$ 
amount to 10-20\% over the full range of $Q^2$. 
We also stress that the present ``Bethe-Salpeter" results, contrary to those 
sometimes referred to in the lite\-ra\-ture under the same name 
\cite{Merten:2002nz}, do not involve any
approximation like an instantaneous one for the interaction. They are fully
relativistic, verifying expected properties under both 
Lorentz transformations and space-time translations. 
Moreover, they satisfy current conservation when applicable. 

To a large extent, the predictions of the above models play the role 
of measurements. In comparison with a real physical problem, the comparison of 
these ``measurements'' with results obtained in the frame of relativistic 
quantum mechanics offers many advantages. The physics is quite simple 
(one-boson exchange, no crossed diagram). Intrinsic form factors, if any, 
cancel in the comparison. Moreover, there is no spin complication. The 
comparison can thus be particularly useful to test minimal ingredients 
entering relativistic quantum mechanics. In spite of this simplicity, 
reproducing the predictions for the asymptotic behavior and especially 
the log terms is not trivial. They therefore represent a severe test for 
RQM approaches, both for the mass operator and the currents. 
They could be quite relevant when looking at  a more realistic problem 
like the pion charge form factor. 
As a side remark, the comparison can tell about different effects which 
have been mentioned in the literature in relation with the implementation 
of relativity. The first one, which is of direct relevance for the present 
work,  concerns the  ``point-form'' approach at high $Q^2$. It was found 
that the dependence of form factors on $Q^2$ was affected by an extra factor 
as follows \cite{Allen:2000ge}: 
\begin{equation}
Q^2 \rightarrow Q^2 \;\Big(1+\frac{Q^2}{4\,M^2}\Big)\,.\label{asym2}
\end{equation}
It provides an asymptotic dependence $Q^{-2n}$ where other approaches 
would give $Q^{-n}$, hence a faster drop off. The second effect concerns 
Lorentz-contraction. In order to take it into account, it was proposed 
to modify  non-relativistic form factors by changing the argument 
as follows: $Q^2 \rightarrow Q^2/ (1+Q^2/(4M^2))$, which leads 
to constant form factors at high $Q^2$.  Amazingly, this recipe involves 
the same relativistic correction factor, $(1+Q^2/(4M^2))$, as in 
Eq. (\ref{asym2}), but at the denominator instead of the numerator. 
It has been however mentioned that the recipe was incorrect. Some analysis 
of what it is missing has been described in Ref. \cite{Amghar:2002jx}.

\section{Form factors: results}

\begin{figure}[htb]
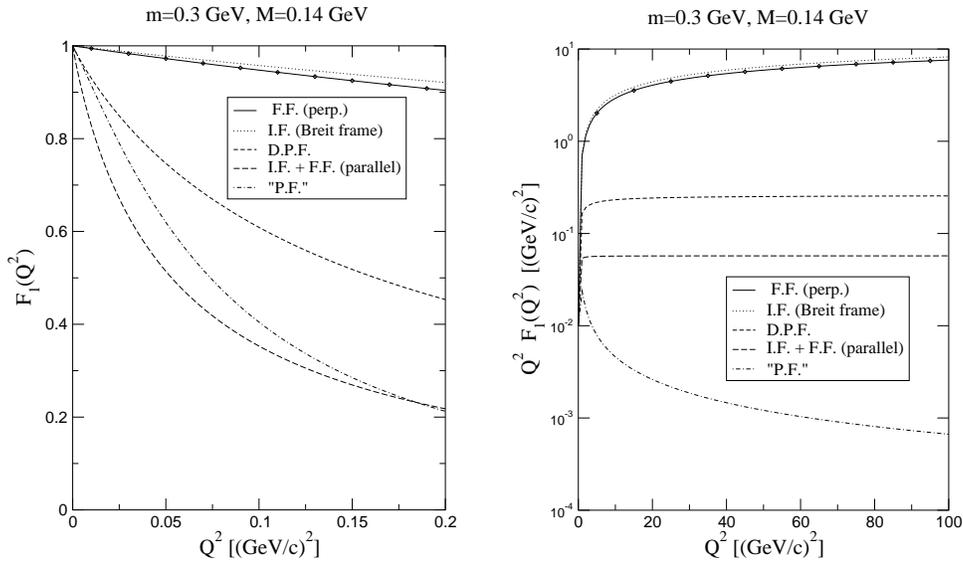

\begin{center}
\mbox{ \epsfig{ file=fraM1s.eps, width=6cm}}
\hspace{3mm}
\mbox{ \epsfig{ file=fraM1S.eps, width=6cm}}
\end{center}
\caption{Charge form factor in various forms of relativistic quantum 
mechanics and an infinite-mass boson-exchange model: left for low $Q^2$ 
and right for high $Q^2$ (the last one is multiplied by a factor $Q^2$ to 
compensate an expected $Q^{-2}$ behavior). The ``exact'' results (our
``experiment'') are represented by diamonds. \label{fig3} }
\end{figure}  
In looking at form factors, two domains are of special interest. At low $Q^2$,
they are sensitive to the radius (charge or else depending on the probe) while 
at high $Q^2$, they ge\-ne\-rally evidence power-law behaviors that can be compared 
to expectations. We therefore present accordingly the results in two figures
for each form factor, up to $Q^2=0.2\,({\rm GeV/c})^2$ and  
$Q^2=100\,({\rm GeV/c})^2$. Moreover, in this second case, we 
show the form factor multiplied by the inverse of its expected asymptotic 
behavior, respectively $Q^2$ and $Q^4$ for the infinite-mass  and zero-mass 
boson exchange. Thus, these last results should evidence a plateau at high 
$Q^2$, up to possible ${\rm log}^2(Q)$ corrections.
\begin{figure}[htb]
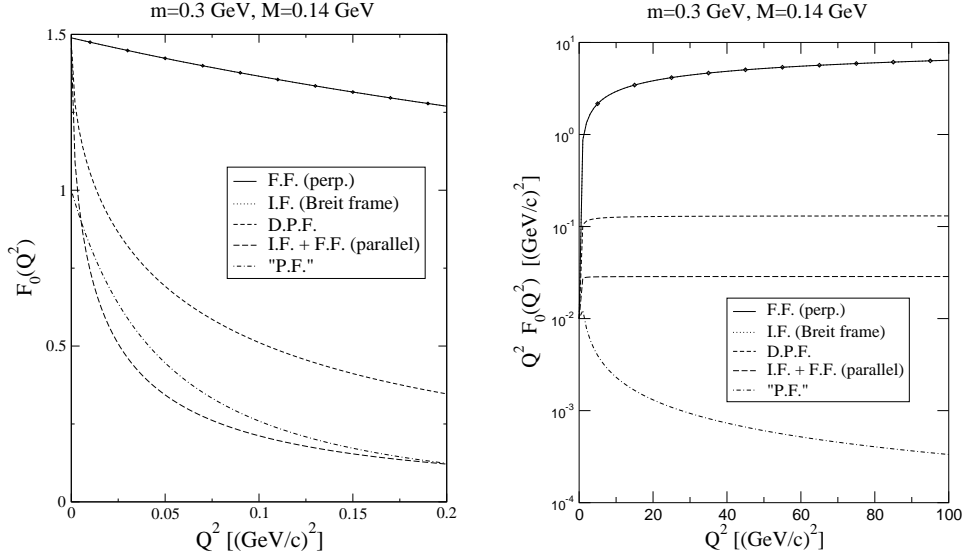

\begin{center}
\mbox{ \psfig{ file=fraM0s.eps, width=6cm}}
\hspace{3mm}
\mbox{ \psfig{ file=fraM0S.eps, width=6cm}}
\end{center}
\caption{Same as in Fig. \ref{fig3} for the scalar form factor.
 \label{fig4} }
\end{figure}  

\begin{figure}[htb]
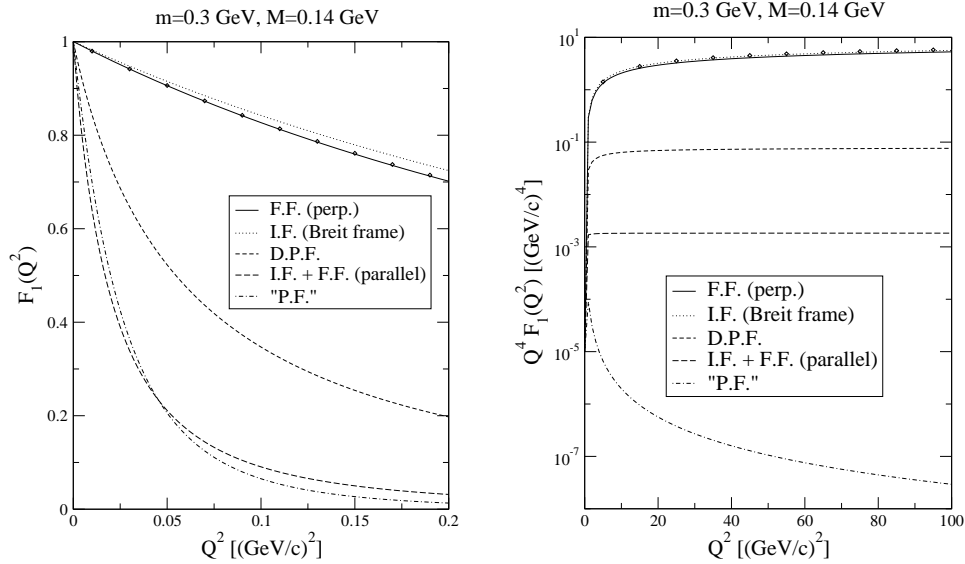

\begin{center}
\mbox{ \psfig{ file=fram1s.eps, width=6cm}}
\hspace{3mm}
\mbox{ \psfig{ file=fram1S.eps, width=6cm}}
\end{center}
\caption{Charge form factor in various forms of relativistic quantum 
mechanics and a zero-mass boson-exchange model: left for low $Q^2$ 
and right for high $Q^2$ (the last one is multiplied by a factor $Q^4$ to 
compensate an expected $Q^{-4}$ behavior). \label{fig5} }
\end{figure}

Form factors for the infinite-mass boson exchange case, $F_1(Q^2)$ and 
$F_0(Q^2)$, are presented in Figs. \ref{fig3} and \ref{fig4} respectively,
while those for the zero-mass boson exchange are shown in Figs. 
\ref{fig5} and \ref{fig6}. Beside the ``exact'' result (our ``experiment'') 
represented by data points, each figure contains six curves:\\ 
- the front-form form factor in the ``perpendicular'' momentum configuration 
$q^+=0$ (F.F. (perp.)),\\
- the instant-form form factor in the Breit frame (I.F. (Breit frame)),\\
- the front-form form factor in a ``parallel'' momentum configuration 
$\vec{q}\parallel (\vec{P}_i+\vec{P}_f)\parallel \vec{n}$ (F.F. (parallel)),\\
- the instant-form form factor in a ``parallel'' momentum configuration $\vec{q}\parallel 
(\vec{P}_i+\vec{P}_f)$ with $|\vec{P}_i+\vec{P}_f| \rightarrow \infty$ which 
coincides with the previous one (I.F. (parallel)),\\
- the form factor calculated in a Dirac's inspired point form (D.P.F.),\\
- and the form factor calculated in an earlier implementation of the  
point-form (``P.F.'').\\
\begin{figure}[htb]
\begin{center}
\mbox{ \psfig{ file=fram0s.eps, width=6cm}}
\hspace{3mm}
\mbox{ \psfig{ file=fram0S.eps, width=6cm}}
\end{center}
\caption{Same as in Fig. \ref{fig5} for the scalar form factor.
 \label{fig6} }
\end{figure}

Examination of Figs. \ref{fig3}-\ref{fig6} shows that the various curves 
clearly fall into two sets, those that are close to the ``experiment'' 
and the other ones that are far apart. This occurs both at low and high $Q^2$. 
The first set comprises standard instant- and front-form calculations while the 
second one includes the same approaches with non-standard momentum
configurations as well as 
point-form results that stem from a Lorentz covariant approach. Looking for an 
argument that can discriminate between different curves, we notice that the 
first set corresponds to approaches where the boost implementation is 
essentially independent of the mass of the system\footnote{Instant form 
in the Breit frame and more generally with $E_i=E_f$ 
\big($\vec{q} \;(=\vec{P}_f-\vec{P}_i) \perp (\vec{P}_i+\vec{P}_f)$\big), 
front form with $E_f-E_i-(\vec{P}_f-\vec{P}_i)\cdot \vec{n}=0 $.} while 
form factors in the second set all depend on $Q$ though the ratio $Q/(2M)$. 
As noticed elsewhere, this feature has the consequence that the charge 
(or Lorentz-scalar)   
squared radius scales like $1/M^2$, a feature which has so surprising effects 
that one can suspect that the underlying formalism misses an important property. 
Of course, this can be repaired by appropriate two-body currents. 

A closer examination at Figs. \ref{fig3}-\ref{fig4}, which correspond 
to an interaction model with an infinite-mass boson exchange, shows 
that some results coincide with ``experiment''. These results include 
the standard front-form results and some instant-form ones. The agreement 
in the first case is not totally surprising. On the one hand, 
the ``zero-range'' nature of the interaction does not leave much 
freedom on the solution of the mass operator. On the other hand, 
in a field-theory approach, it is known that corrections due to Z-type 
diagrams are suppressed in the case of scalar particles and $q^+=0$ 
\cite{Karmanov:1991fv}. This result cannot be applied to relativistic 
quantum mechanics but, taking into account that a Z-type diagram 
has often a contact term as a counterpart in this formalism, 
it makes the absence of correction in this case plausible. The agreement 
in the instant form for the scalar form factor is more surprising. 
As mentioned in Sect. 2, the current was including a correction ensuring 
that $F_0(Q^2=0)$ be Lorentz invariant (factor $cf_0$ in Eq. (\ref{ff3})). 
It turns out that this constraint entails the identity of the form factor 
$F_0(Q^2)$  with the ``experiment'' at all  $Q^2$. In the case of  
the charge form factor, $F_1(Q^2=0)$ was in any case Lorentz invariant, 
requiring no correction. We notice however that a correction to the current 
could have been introduced in Eq. (\ref{ff1}), preserving the Lorentz 
invariance of $F_1(Q^2=0)$, while providing identity with 
the ``experiment'' \cite{Amghar:2002jx}. 

A last remark about the infinite-mass boson results concerns 
the asymptotic behavior of form factors. Most of them scale like 
$Q^{-2}$. This is clearly seen for some of the results but not so clear 
for the ``experiment'' and the standard instant- and front-form results. 
In these cases, the slight increase (after multiplying form factors 
by $Q^2$) is due to non-trivial log$^2(Q)$ corrections whose reproduction 
is a stringent test of the implementation of relativity. The  only 
exception concerns the ``point-form'' implementation that produces 
form factors scaling like $Q^{-4}$. This behavior can be traced back to the 
observation that results in Eq. (\ref{asym2}). As this feature is not 
evidenced by the Dirac's point-form, we believe that it is a specific 
feature of the approach.  Most probably, it is due to the fact that, 
contrary to all other ones, it implies different surfaces in the 
description of initial and final states. 

The general pattern of results presented in Figs. 5, 6, which correspond 
to a zero-mass boson exchange, is similar to that of   Figs. 3, 4. 
Significant differences are nevertheless worthwhile to be mentioned. 
Considering first the ``good'' results, those for the standard instant and 
front forms are close to the ``experiment'' but none is identical. 
Due to the long range of the 
underlying interaction model, some uncertainty is expected in the 
derivation of the mass operator. We however notice that the ``good'' 
result is largely due to the choice of the high momentum behavior 
of the interaction $V(\vec{k},\vec{k}')$ entering the mass operator. 
As noticed by Alabiso and Schierholz \cite{Alabiso:1974sg}, the behavior 
of form factors at high $Q^2$ is closely related to the interaction 
one at high $k$, which has thus to fulfill well determined conditions. 
This result is important as it provides minimal guidelines 
for further work concerning the pion form factor for instance. 
At low $Q^2$, one can notice some discrepancy for the form factor $F_0(Q^2)$,
which, contrary to the charge form factor, $F_1(Q^2)$, is not protected 
by some charge conservation. This points to the missing contribution of
relatively standard two-body currents. It is nevertheless interesting that the
two-body currents implied by the introduction of the factor, $cf_0$, in Eq.
(\ref{ff3}), make the standard instant- and front-form ones equal to each other
at $Q^2=0$, while decreasing their difference at non-zero values of $Q^2$. The
smaller discrepancy, which does not exceed 6\% (instead of 30\%) points to a
partial restoration of Lorentz invariance. 
Considering now the ``bad'' results, it is found that  the sensitivity 
to the approach under consideration is much larger than for 
the zero-range interaction model. This increased sensitivity is 
in relation with the expected asymptotic behavior $Q^{-4}$, which 
implies a ratio of instant- and front-form results with standard and 
non-standard momentum configurations of the order $(2\,m/M)^4$, instead 
of $(2\,m/M)^2$ previously (up to log terms). As for the ``point-form'' 
approach, the asymptotic behavior is $Q^{-8}$, as expected from the above 
$Q^{-4}$ behavior together with the modification given by  Eq. (\ref{asym2}). 

A few comments have already been made about the results obtained in a  Dirac
motivated point-form. As such results are presented for the first time,
it is appropriate to discuss them separately a little more. 
It is first noticed that its Lorentz covariance does not ensure 
it provides ``good'' results. While it does better than 
an earlier ``point-form'' approach, especially with respect
to the asymptotic behavior of form factors that now evidence the right $Q^2$ 
power law, it suffers from the fact that their dependence on the momentum
transfer involves the ratio $Q/2M$, implying obvious drawbacks in the limit 
$M \rightarrow 0$. Actually, it turns out that these undesirable features 
are shared by instant- and front-form approaches with unusual momentum
configurations, which suggests that the problems raised in the above limit 
have a more general character, independent of the intrinsic 
Lorentz covariance of the point-form approach. A second observation 
concerns how these point-form results
compare quantitatively with other ones. Representing a weighted average 
of contributions that involve in particular the standard and non-standard 
front-form approaches, it is not surprising that the new point-form results 
fall in between. Correcting the misleading impression produced by 
the logarithmic scale in the right pa\-nels of Figs. \ref{fig3}-\ref{fig6}, 
these results are however closer to the later ones (non-standard) 
than to the former ones (standard momentum configuration). 
Throughout this paper, 
we consi\-de\-red a strongly bound system. Apart from the fact that some
of the inputs correspond to a physical system (the pion), an extreme case like
the one we considered offers the advantage of better emphasizing the 
peculiarities pertinent to the formalism. Looking at a weakly bound system 
would not have been so instructive. 

Results for form factors presented in this section show that they strongly
depend on the underlying formalism when only the single-particle current is
considered. Though no detailed study was made here, it appears that the largest
discrepancies with ``experiment" can be interpreted as if the momentum
transfer was effectively larger than the physical one. The factor could be 
of the order $2\,\bar{e_k}/M$ in the instant and front forms with ``pa\-ral\-lel''
momentum configuration as well as in the Dirac's inspired point form ($\bar{e_k}$ 
is some average value for the internal kinetic energy). An extra factor, 
$(1+Q^2/4M^2)^{1/2}$ should be considered for the ``point form". 
The discrepancy between the form factors calculated in the standard 
instant- and front-form approaches does not exceed a few percent's. 
This roughly summarizes the main features of numerical results presented 
in this section.

\section{Discussion and relationship to Poincar\'e 
space-time translation invariance}
In view of the results presented in the previous section, the question 
arises of whether there is a way to discriminate results from a simple 
argument and, possibly, to remove the main discrepancies. We first 
notice that a Lorentz-covariant approach like the point form, 
which {\it a priori} represents a sensible feature, does not guarantee 
to get a ``good'' result. Such situations often occur in physics. 
In a region of intrinsically deformed nuclei for instance, 
the binding energy of a spherical nucleus ($J=0$) is better obtained 
by using a  mean field which breaks the spherical symmetry. 
Observing that the ``good'' results are obtained in the following cases, 
instant form with  $E_f=E_i$ (this goes beyond the standard Breit frame 
mostly mentioned in the present work) and front form with 
$\xi \!\cdot\! (P_i-P_f)=q^+=0$, a more important criterion 
could be the conservation of the 4-momentum at the interaction vertex 
of the external probe with the struck constituent. This condition, 
which is fulfilled in field-theory models, can only be verified 
approximately in relativistic quantum mechanics. The best that 
one can require is that this condition be fulfilled on the average, 
$<(p_i+q-p_f)^{\mu}>=0$ (which is a much weaker constraint than 
$ (p_i+q-p_f)^{\mu}=0$). As the conservation of the 4-momentum 
stems from Poincar\'e space-time translation invariance, one can 
also infer that an equivalent criterion involves the conditions 
that make this result possible. With this respect, it was noticed by Coester 
that the momentum $p$ in the point-form kinematics does not generate 
translations consistent with the dynamics \cite{Coester:2003zh}.

Quite generally, Poincar\'e  covariance implies that 
a 4-vector (or a scalar) current transforms under space-time translations as
follows:
\begin{equation}
e^{iP \cdot a}\;J^{\nu}(x) \;(S(x))\;e^{-iP \cdot a}=J^{\nu}(x+a) \;(S(x+a)),
\label{trans}
\end{equation}
where $P^{\mu}$ is the 4-momentum operator of the Poincar\'e algebra that
generates space-time translations. In a particular case, this equation reads: 
\begin{eqnarray}
J^{\nu}(x) \;(S(x))=e^{iP \cdot x} \; J^{\nu}(0) \;(S(0))\;e^{-iP \cdot x}.
\label{transl2}
\end{eqnarray}
Matrix elements of this  relation between states 
with 4-momentum $P^{\mu}_i$ and $P^{\mu}_f$ can be considered. 
By construction of the Poincar\'e algebra pertinent to a RQM approach, 
these states are eigenstates 
of the 4-momentum  $P^{\mu}$ with eigenvalues $P^{\mu}_i$ and $P^{\mu}_f$. 
This allows one to factorize the $x$ dependence of the matrix elements 
as ${\rm exp}(i\,(P_i-P_f)\cdot x)$. The integration over $x$ of this factor 
together with the factor  ${\rm exp}(i\,q\cdot x)$ describing the external 
probe then provides the well known energy-momentum conservation relation 
$ (P_i+q-P_f)^{\mu}=0$.
Quite ge\-ne\-rally, fulfilling the above relations requires 
the consideration of many-body components in the current 
$J^{\nu}(x)\; ({\rm or}\;S(x))$, beside the one-body component 
most often retained in the calculations. 
Covariant transformations of currents under  space-time translations 
in RQM approaches can therefore imply further constraints 
beyond the usual energy-momentum conservation relation 
that is made possible by these covariance properties 
and is supposed to hold in any case. 

Further equations that could be more amenable to some check are
obtained by considering an expansion of Eq. (\ref{trans}) 
for small space-time translations. In the simplest case, 
they read \cite{Lev:1993}:
\begin{equation}
\Big[ P^{\mu}\;,\; J^{\nu}(x)\Big]=-i\partial^{\mu}\,J^{\nu}(x),
\;\;\;
\Big[ P^{\mu}\;,\; S(x)\Big]=-i\partial^{\mu}\,S(x)\, .
\label{trinv}
\end{equation}

While some information about the many-body components entering 
the current $J^{\nu}(x)\; ({\rm or}\;S(x))$ can 
be obtained from a parallel study within field theory, the above constraints 
are, in first place, the proper way to introduce 
them in the RQM formalism used here, where they reduce to two-body ones. 
In their absence, one can at least demand that the matrix elements 
of the r.h.s. and l.h.s. of the above equations be equal. For form factors 
considered in the present work, the commutator of the momentum operator 
at the l.h.s. can be transformed into a matrix element involving the
momentum transfer $q^{\mu}$, using the conservation of the overall momentum.
The equality requirement with the r.h.s. then implies that the momentum 
transferred  to the system of interest and that one transferred 
to the struck constituent be the same on the average, 
$<q^{\mu}>=<(p_f-p_i)^{\mu}>$. 
This is precisely the condition that is suggested for getting ``good'' 
form factors. Looking at the other results, it appears that they 
all violate this equality. In the instant form with the ``parallel'' 
momentum configuration, 
the equality is verified for the spatial components but not for the time one. 
In the point form, due to the presence of the interaction in the momentum, 
it is {\it a priori} violated in all components. 

The discussion can be extended to any number of commutators or 
derivatives in Eqs. (\ref{trinv}). Among them, the relations involving 
the following double commutators: 
\begin{equation}
\Big[ P_{\mu}\;,\Big[ P^{\mu}\;,\; J^{\nu}(x) \;({\rm or}\;S(x))\Big]\Big]=
-\partial_{\mu}\,\partial^{\mu}\,J^{\nu}(x)\;({\rm or}\;S(x)) \,,
\label{doublec}
\end{equation}
and its matrix elements at $x=0$:
\begin{equation}
<\;|q^2\; J^{\nu}(0) \;({\rm or}\;S(0))|\;>=
<\;|(p_i-p_f)^2\,J^{\nu}(0)\;({\rm or}\;S(0))|\;> \,,
\label{doubleme}
\end{equation}
are particularly relevant here. In getting the l.h.s., we made use 
of the energy-momentum conservation. This relation allows one 
to replace by $q^2$ the product $(P_f-P_i)^2$ which results 
from applying the momentum operator in Eq. (\ref{doublec}) 
on the initial and final states. 
The appearance at the l.h.s. of the scalar quantity, $q^2 (=-Q^2)$, 
on which form factors considered in this work exclusively depend, 
provides a basis for a quantitative discussion 
better than $q^{\mu}$. As for the quantity, $(p_i-p_f)^2$, appearing 
at the r.h.s. of the last equation, we notice at this point that the use 
of Eq. (\ref{pisupf}) allows one to write it as:
\begin{eqnarray}
(p_i-p_f)^2&=&\Big((P_i-P_f)^{\mu} 
+\xi^{\mu}\;(\Delta_i -\Delta_f)\Big)^2
\nonumber \\
&=&(P_i-P_f)^2+2\,\xi \!\cdot\! (P_i-P_f)\;(\Delta_i -\Delta_f) 
+\xi^2\;(\Delta_i -\Delta_f)^2 \,,
\label{doubled}
\end{eqnarray}
where
\begin{equation}
\Delta_{(i,f)}=\frac{4\,e^2_{k_{(i,f)}}-M^2}{
\sqrt{ (\xi \!\cdot\! P_{(i,f)})^2 + (4\,e^2_{k_{(i,f)}}\!-\!M^2)\;\xi^2 } 
+  \xi \!\cdot\! P_{(i,f)} }\,. 
\label{q2}
\end{equation}
As the examination of the last quantity shows, the discrepancy with 
the quantity $q^2$ appearing at the l.h.s of Eq. (\ref{doubleme})  
involves interaction effects. We stress that the factor  $(p_i-p_f)^2$ 
depends on derivatives of the current around $x=0$, implying therefore 
current properties going beyond those required to obtain the
energy-momentum conservation relation.

We now compare the matrix elements of both sides of Eq. (\ref{doublec}), 
numerically, or analy\-ti\-cally when possible, while retaining only the 
one-body component of the current as done most often. 
Beginning with the standard front-form approach 
($\xi^2=0,\; \xi \cdot (P_i-P_f)=0$), one can easily check from the
expression of $(p_i-p_f)^2$, Eq. (\ref{doubled}), that the equality
is always fulfilled. In the standard instant-form approach ($E_i=E_f$), 
the second term at the r.h.s. of Eq. (\ref{doubled}) vanishes and only 
the last term ($\propto \xi^2$) provides some departure. 
This one is found to amount to 20\% at most (at low $Q^2$) and vanishes 
when the average momentum of the system goes away from the Breit-frame 
case, $\vec{P}_i+\vec{P}_f=0$, to the limit 
$|\vec{P}_i+\vec{P}_f| = \infty$, where the above front-form 
result is recovered. 

Contrary to the above results, those for the instant and front forms 
in the ``parallel" momentum configuration involve a non-zero contribution 
from the second 
term at the r.h.s. of Eq. (\ref{doubled}) but none from the last term. 
For the charge form factor, the departure to the expected equality 
of the two members of Eq. (\ref{doubleme}) decreases
from a factor 45 at low $Q^2$ to 30 at the highest values of  $Q^2$ considered 
in this work. Typically, this factor is of the order $(2\,\bar{e_k}/M)^2$, 
as suggested by Eq. (\ref{doubled}). In the Dirac's point-form inspired 
approach, the departure, which is also produced by the second term 
at the r.h.s. of Eq. (\ref{doubled}), decreases from a factor 15 to 3, 
a value which is intermediate between the two front-form ones given above. 
Finally, for the ``point-form'' approach, where one has to account 
for different $\xi^{\mu}$ in the initial and final states, the departure 
is found to increase from a factor 30 at low $Q^2$ to about 35000 
at $Q^2=100\,({\rm GeV/c})^2$, the factor being roughly given 
by $((2\,\bar{e_k}/M)\; (E_{Q/2}/M))^2$, as can be checked from Eq.
(\ref{doubled}). To a large extent, the above departures 
are in accordance with those inferred from the numerical examination 
of form factors in the previous section. This shows that 
the discrepancy of the ``bad'' form factors with the ``exact'' ones 
is closely related to a violation of properties related to 
Poincar\'e translational covariance of currents. 
In comparison to the large effects mentioned above, we notice 
that the violation of Lorentz invariance is only a few 10\% in cases 
where the above properties are approximately
fulfilled (see effect for the scalar form factor $F_0(Q^2=0)$ or that one
resulting from the difference of the standard instant and front-form form
factors at any $Q^2$).

As can be expected, the above departures point to the missing contribution 
of interaction effects. This is easily checked in the most striking cases 
where the factor multiplying $q^2$ at the r.h.s. of Eq. (\ref{doubleme}) 
is given by $4\,e^2_k/M^2$. As the consideration of various examples
shows \cite{Desplanques:2003nk}, the numerator has often to be completed 
by an interaction term as follows, $4\,e^2_k \rightarrow
(4\,e^2_k+4\,m\,\tilde{V})$. When this is done, the use of the mass 
equation, Eq. (\ref{wf6}), allows one to replace 
 $(4\,e^2_k+4\,m\,\tilde{V})/M^2$ by 1. It is not rare that such a result 
stems from the restoration of some symmetry properties, in relation with  
Poincar\'e translational covariance of currents in the present case. 
In principle, the missing interaction effects should be accounted 
for by two-body currents pertinent to the underlying formalism 
so that to ensure the equivalence of different approaches. 

Examining the problem of these two-body currents, we found that they 
differ from those mostly encountered in nuclear physics for instance. 
In some limit, they have the form $0/0$, which rather suggests 
that their role is related to the restoration of a symmetry. Only such 
currents, with a non-trivial behavior, can correct either for the paradox 
of a radius going to $\infty$ while the mass of the system goes to 0  
(which can be obtained by increasing the attraction), or for the discrepancy 
with dispersion approaches (which cannot {\it a priori} produce such 
a radius scaling). Actually, these currents involve a slowly converging series 
in terms of the coupling constant, generally requiring an infinite number 
of contributions tending individually to zero in the $M \rightarrow 0$ 
limit.  Having a specific role, it is not a surprise 
if such currents could not be obtained in an earlier work where 
two-body currents were motivated by current conservation and the high-$Q^2$ 
behavior of the Born amplitude \cite{Desplanques:2003nk}. 
Evidently, these conventional currents can play a role but rather 
at the level of the relatively slight difference between the standard 
instant- and front-form form factors. This is illustrated by results 
presented in the previous section for the scalar form factor where 
this difference was reduced, therefore tending to restore Lorentz invariance 
as far as these approaches are concerned.

\begin{figure}[htb]
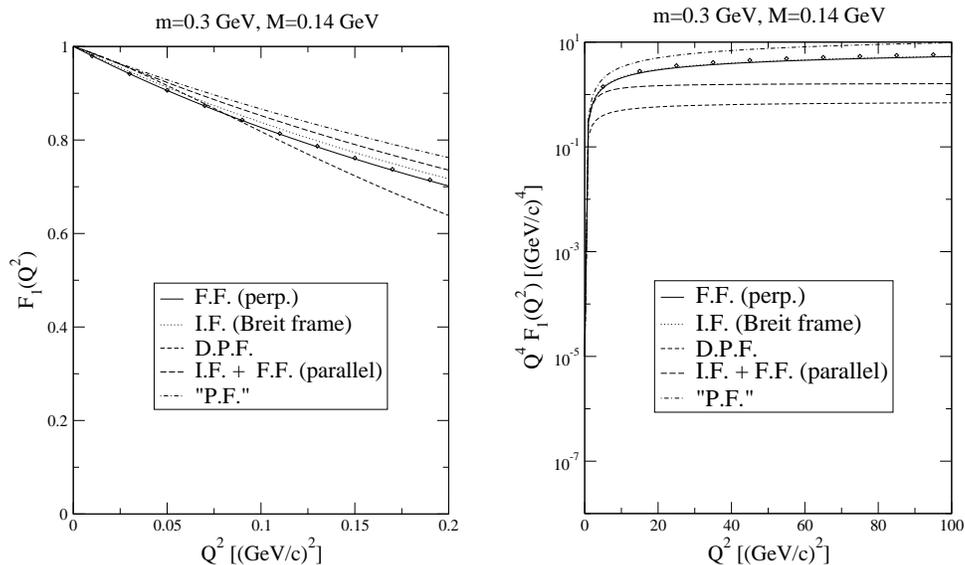

\begin{center}
\mbox{ \epsfig{ file=fram1sr.eps, width=6cm}}
\hspace{3mm}
\mbox{ \epsfig{ file=fram1Sr.eps, width=6cm}}
\end{center}
\caption{Charge form factor in various forms of relativistic quantum 
mechanics and a zero-mass boson-exchange model: same as in Fig. \ref{fig5} 
but with including contributions that partly correct for missing properties
related to Poincar\'e translational covariance of currents. 
\label{fig7} }
\end{figure}

An alternative but approximate way to account for the missing interaction
effects is suggested by the above observation about their role 
in restoring some symmetry. To remove the effect of the undesirable factor,   
$4\,e^2_k/M^2$, it suffices to compensate for it in the matrix element 
of the single-particle current. This can be done by multiplying 
the coefficient of the squared momentum transfer, $Q^2$, 
appearing in the calculations, by the inverse of the above factor. 
This represents a schematic way 
to proceed. In practice, there are some corrections to this factor, 
especially for the ``point-form" approach. An improved implementation 
of the missing interaction effects  along the above lines thus consists 
in introducing, in place of $4\,e^2_k/M^2$, the departure factor
resulting from the numerical comparison of both sides of Eq. (\ref{doubleme}). 
This can be done analytically in the standard instant-form  and 
in the ``point-form" approaches. Equation (\ref{doubleme}) is then 
fulfilled exactly (which does not necessarily imply that 
symmetry properties related to the translational covariance of currents
are fully restored). In the other cases, it is  done numerically. 
No change is required for the standard front-form approach which 
satisfies this equation identically as already mentioned. 

The corrected results are presented in Fig. \ref{fig7} for the charge 
form factor calculated in the infinite-range interaction model at both 
low and high $Q^2$. Comparison with the corresponding results presented 
in Fig. \ref{fig5} shows  a spectacular decrease of the discrepancies.
At low $Q^2$, the slope of the form factor, which determines 
the charge radius, differs from the standard front-form one by 25\% 
at most instead of a factor up to 20 before. At large $Q^2$, 
the discrepancy now reaches one order of magnitude instead 
of a few  orders, up to 7 in the worse case. Altogether, these results
demonstrate that properties related to Poincar\'e space-time translations 
are relevant in describing form factors reliably. 
These properties go beyond the energy-momentum conservation generally 
expected from the Lagrangian invariance under these transformations.
At the same time, 
all the drawbacks that appear especially in the limit of a zero-mass 
system and could point to the violation of some symmetry are removed. 
Somewhat incidentally, the above approach was used  
in Ref. \cite{Allen:1998hb}, allowing the authors to get a reasonable 
result for the pion charge form factor, but no justification could 
be given within the underlying formalism. It was also noticed 
that the same wave function could give rise to quite different form factors, 
depending on the total mass of the system \cite{Desplanques:2003nk}. 
This arbitrariness is largely removed by fulfilling the  constraints 
discussed in this section.

There remain some discrepancies. In this respect, we notice that 
the above method works relatively better when the violation 
of the equality given by Eq. (\ref{doubleme}) is larger. Thus, 
for the case where the remaining discrepancy for the form factor is 
the largest (D.P.F. at large $Q^2$), the violation 
of Eq. (\ref{doubleme}) was of the order of a factor 3 before correction. 
This indicates that some refinement is necessary\footnote{It can be shown 
that predictions for the charge form factor in different forms 
as well as in a dispersion-relation approach \cite{Krutov:2002}(
partly corrected) could be made identical \cite{Desplanques:2007}. 
This result supposes a single-particle-like current slightly different 
from what is generally assumed}. 
Pursuing along this line in the cases where the violation 
of Eq. (\ref{doubleme}) is originally large is questionable however. 
Using an approach that introduces interaction effects that turn out to be
fictitious, as they have to be removed later on in one way or another, 
is not the best strategy, especially if further corrections have to be
considered. Thus, from the present study, only the standard front-form 
approach ($\xi\cdot(P_i-P_f)=q^+=0$) or the the instant-form one 
($\xi\cdot(P_i-P_f)=E_i-E_f=0$) appear as viable when the current 
is restricted to a single-body one. The last approach 
includes the standard Breit-frame case ($\vec{P}_i+\vec{P}_f=0$) 
but also the case $\vec{P}_i+\vec{P}_f\neq 0$ with 
$(\vec{P}_i+\vec{P}_f) \perp (\vec{P}_i-\vec{P}_f)$, which allows one 
to make the relation with the previous front-form case in the limit 
$|\vec{P}_i+\vec{P}_f| \rightarrow \infty$. It has been often thought that 
these frameworks were more relevant than other ones. The fact that they 
better fulfill constraints from translational covariance of currents 
provides a sensible justification. 
Most likely, the argument based on these constraints can be extended 
to field-theory type calculations made in the light front 
\cite{Simula:2002vm,Bakker:2000pk,deMelo:2002yq}. 
In these ones, the condition $q^+=0$ is currently used but does not seem 
to have received any justification other than providing results 
close to the exact ones \cite{Karmanov:1991fv}.


\section{Conclusion}
In this paper, we presented charge and scalar form factors corresponding to 
different forms or different momentum configurations, calculated using 
a single-particle current. This has been done for two opposite extreme 
interaction models corresponding to the exchange of an infinite- 
and a zero-mass boson. In comparison to a previous work \cite{Amghar:2002jx}, 
the present one includes results for instant and front forms 
in a ``parallel'' momentum configuration, which are most often ignored, 
and a point form inspired from Dirac's work, which differs 
from the currently referred one \cite{Sokolov:1985jv,Bakamjian:1961}. 
A method allowing one to restore Lorentz invariance 
for the scalar form factor at $Q^2=0$ is presented. 
Anticipating on a future work, the constituent mass and the mass 
of the system are those appropriate for the pion. 
As noticed elsewhere, the small mass of the system in comparison 
of the constituent one, apart from the fact it corresponds 
to a physical case, makes differences between various 
approaches more striking.  

At low $Q^2$, the results clearly fall into two sets, close and sometimes 
identical to the ``experiment'' for some of them, far apart for the other ones. 
In the first category, one finds the standard instant- and front-form results 
while the second one contains  the instant- and front-form results for 
non-standard momentum configurations and the point-form ones. 
The quantity characterizing 
these different results is the squared charge or scalar radius. Up to some 
numerical factor, it is roughly given by the inverse of the  squared 
constituent mass in one case (notice that the binding is here close 
to the sum of the constituent masses) and by the  inverse of the 
squared mass of the system in the other. This last property results 
from the dependence of form factors on $Q^2$ through the ratio $Q^2/(2M)^2$. 
At high $Q^2$, a similar separation into two sets occurs. 
Again, the mass of the system is the essential parameter 
\cite{Desplanques:2001ze,Desplanques:2001zw,Amghar:2003tx,Coester:2004qu}. 
One can guess however that the ``point-form'' behaves differently. 
As noticed in Ref. \cite{Allen:2000ge}, the asymptotic behavior 
of the corresponding form factors is rather $(Q^{-n})^2$ where other 
approaches give $(Q^{-n})$. 

Altogether, there is not much doubt about which approach is 
an efficient one in the sense that most of the contribution is due 
to a single-particle current.  The other ones (ins\-tant and front forms 
with $\xi\cdot(P_i-P_f)\neq 0$ and point forms) require large if not 
dominant contributions from two-body currents, which is not the best for 
incorporating further corrections or for discussing physics when a comparison 
to experiment is done. Interestingly, the relevant values  of the 
spectator-particle momentum in these approaches are consi\-de\-rably increased 
in comparison to the value of the order of $Q$ in the case where ``good''
results are obtained. The enhancement, which partly depends 
on the momentum confi\-gu\-rations,
currently involves  a factor of the order $2\,m/M$. In the ``point form''
approach, an extra enhancement factor,  $1+Q^2/(2M)^2$, has to be considered 
for the struck particle. These approaches, especially the last one, 
therefore require a significantly improved implementation 
of relativity. This qualitatively agrees with the expected 
increased role of two-body currents mentioned above. 
The important point we want to stress however 
is that the earlier ``point-form'' approach is not any more the only 
one to evidence obvious drawbacks due to the dependence of form factors 
on $Q^2$ through the ratio $Q^2/(2M)^2$. Actually, we are inclined to 
believe that the ``good'' result is an exception while the ``bad'' result 
is more likely the rule as soon as calculations are made 
for an arbitrary momentum configuration. 

Looking for an argument that could explain the discrepancy evidenced 
by the compa\-ri\-son of different approaches in a single-particle current 
approximation, we found a close relationship between two features. 
The first one involves the departure of the form factors to the ``exact" 
results (or the standard front-form ones). The second one is concerned with 
the violation of an equality relating the squared-momentum transferred 
to the whole system and that one transferred to the constituents, 
Eq. (\ref{doubleme}),
which stems from Poincar\'e space-time translation covariance. 
Absent in the standard front-form case, the violation amounts to 20\% 
at most in the instant-form one. In the other cases, where the effect 
is more striking and can reach orders of magnitude, the violation 
is roughly given by a factor $4\,e^2_k/M^2$ possibly corrected by a factor 
$1+Q^2/(4M^2)$ in the ``point-form" case. Again, the violation points 
to the missing contribution of two-body currents. 

Examining 
the explicit derivation of these two-body currents, we find it is rather
hopeless as it requires the consideration of terms to all orders in the
interaction in the limit of a zero-mass system. Another approach is suggested 
by the expression of the above discre\-pan\-cy factor, $4\,e^2_k/M^2$, which
deviates from the expected value 1 by a term which involves the interaction. 
It consists in rescaling the factor multiplying $Q^2$ in the expression 
of the form factor so that the constraints from transformations 
of the current under Poincar\'e space-time translations discussed above 
be fulfilled. This method, which, in some sense, amounts to sum up some 
of the above two-body 
contributions, has been applied to form factors considered in this work. 
It is found to remove the largest discrepancies that some of the approaches 
were evidencing either with ``experiment" or with the standard front-form
results that are close to it. In particular, all the drawbacks related to a
small ratio of the total mass to the kinetic energy, $M/(2\,e_k)$, vanish. 
We believe that these results, which involve many orders of magnitude effects,
demonstrate the relevance of properties related to Poincar\'e space-time 
translations, beyond the usual energy-momentum conservation. 
In comparison, effects due to a violation of Lorentz
invariance evidenced in the present work, assuming that the other symmetry
approximately holds, only amount to a few 10\%. In this respect, 
we notice that  large effects attributed to a violation of the first 
symmetry, like a frame dependence, could be actually due to a
simultaneous violation of the second one.

Part of this work was originally motivated by the drawbacks evidenced 
by a ``point-form'' approach for the calculation of form factors. 
An implementation more in the spirit 
of the Dirac's point-form has not alleviated much the problems. While 
these approaches fulfill Lorentz covariance, it sounds from the present 
work that a more sensible criterion might be Poincar\'e space-time 
translation covariance. With this respect, the Lorentz invariance of the
form factors in the point-form approach, which represents {\it a priori} 
a desirable property, turns out to be a disadvantage as it implies 
that there is no frame where the effect of the violation of the other 
symmetry can be minimized. Different aspects of Poincar\'e covariance 
have been discussed at length in the past, especially in relation 
with rotation symmetry within the front-form dynamics (see for instance 
Ref. \cite{Lev:1998qz}). The relevance of translational-covariance properties 
for applications of relativistic quantum mechanics  
has hardly been discussed however. It should probably be given 
increased attention  in the future. In the case of form factors, 
this should be facilitated by the test and the subsequent correction 
we proposed. In short, the squared momentum transferred to the constituents
should match as much as possible that one transferred to the system 
under consideration, which, after all, is not surprising.

\noindent
{\bf Acknowledgements}\\
We are very grateful to T. Melde for a question about a possible relation 
of the strange behavior of form factors in some approaches and 
a possible violation of space-time translation invariance, which our results 
were tending to support. The question motivated further search for 
a more quantitative argument, which is presented in the last part of the paper.

\appendix

\section{Derivation of form factors in ``parallel'' momentum configurations 
and $\bar{P} \rightarrow \infty$}
\label{app:a}
We here derive successively expressions taken by the instant 
and front-form form factors in the ``parallel'' momentum configuration, 
$ \vec{P}_i \parallel \vec{P}_f$, 
with the further condition $\bar{P} \rightarrow \infty$ in the former case, 
$ \vec{P}_i \parallel \vec{P}_f \parallel \hat{n}$ in the latter. \\

\subsection {Instant-form case}
Defining the longitudinal direction as the one carried by the common direction 
of $\vec{P}_i$ and $\vec{P}_f$, we introduce the parallel and perpendicular 
components as follows:
\begin{eqnarray}
&&\bar{P}=\frac{1}{2}\;(P_i+P_f)_{\parallel}\,, \hspace{2.5cm} 
(P_i)_{\perp}=(P_f)_{\perp}=0,\nonumber \\
&&Q_{\parallel}=(P_f-P_i)_{\parallel}=2\,y\;\bar{P}, \hspace{1cm}
Q^2_{\parallel}=\frac{Q^2\;(Q^2+4\,(M^2+\bar{P}^2))}{Q^2+4\,M^2}\,, \nonumber \\
&&y_{\bar{P}\rightarrow \infty}
=(\frac{Q_{\parallel}}{2\,\bar{P}})_{\bar{P}\rightarrow \infty}
=\sqrt{\frac{Q^2}{Q^2+4\,M^2}}=v_{B.F.}=v\,,
\nonumber \\
&&p_{\parallel}=x\;\bar{P}, \hspace{0.5cm}
(p_i)_{\parallel}=(1-x)\;\bar{P}-\frac{Q_{\parallel}}{2}\,, \hspace{0.5cm}
(p_f)_{\parallel}=(1-x)\;\bar{P}+\frac{Q_{\parallel}}{2}\,. \label{appa1}
\end{eqnarray}
One has now to consider the limit $\bar{P} \rightarrow \infty$ 
of the different factors entering Eqs. (\ref{ff1}, \ref{ff3}). 
The results are simply listed below for quantities that multiply 
the wave functions in the integral displayed there:
\begin{eqnarray}
&&\Bigg(\frac{d\vec{p}}{e_p}\Bigg)_{\bar{P} \rightarrow \infty}
=d^2k_{\perp} \; \frac{dx}{x}\,,
\nonumber\\
&&\Bigg(
\frac{(e_i+e_p)\;(e_f+e_p)}{4\,e_i\;e_f}\Bigg)_{\bar{P} \rightarrow \infty}
=\frac{1-v^2}{4\,\Big((1-x)^2-v^2\Big)}\,,
\nonumber\\
&&\Bigg(
\frac{2\,(e_i+e_f)}{e_i+e_f+2\,e_p}\Bigg)_{\bar{P} \rightarrow \infty}
=2\,(1-x)\,, 
\nonumber\\
&&\Bigg(e_p \;
\frac{(e_f+e_p)(e_i+e_p)-E_f\;E_i}{(e_f+e_p)(e_i+e_p)(e_f+e_i)}
\Bigg)_{\bar{P} \rightarrow \infty}=0\,.
\label{appa2}
\end{eqnarray}
Dealing with the wave function, $\phi_0(k)$, entering the expression 
of the form factors, Eqs. (\ref{ff1}, \ref{ff3}), is more delicate. Its argument 
can be written in terms of $x$, $k^2_{\perp}$, $\bar{P}$ and $Q_{\parallel}$. 
For the initial state, it reads: 
\begin{eqnarray}
k_i^2=k^2_{\perp}+\frac{1}{2}\;
\Bigg( 
\sqrt{m^2+k^2_{\perp}+\Big((1-x)\;\bar{P}-\frac{Q_{\parallel}}{2}\Big)^2} 
\;\;\sqrt{m^2+k^2_{\perp}+ (x\,\bar{P})^2}
\nonumber\\
-  \Big((1-x)\;\bar{P}-\frac{Q_{\parallel}}{2}\Big)\; (x\,\bar{P}) \Bigg)\,.
\hspace{4cm} \label{appa3}
\end{eqnarray}
Replacing $Q_{\parallel}$ in terms of $y$ and $\bar{P}$, and taking 
the limit $\bar{P} \rightarrow \infty$, one successively gets:
\begin{eqnarray}
(k_i^2)_{\bar{P} \rightarrow \infty}&=&\Bigg(k^2_{\perp}+
\frac{ (m^2+k^2_{\perp})\;
\Big((1-2\,x)\;\bar{P}-\frac{Q_{\parallel}}{2}\Big)^2 }{
4\,x\,\bar{P}\,\Big((1-x)\;\bar{P}-\frac{Q_{\parallel}}{2}\Big) }
\Bigg)_{\bar{P} \rightarrow \infty}
\nonumber\\
&=& \Bigg(k^2_{\perp}+(m^2+k^2_{\perp})\; \frac{(1-2x-y)^2}{4\,x\,(1-x-y)}
\Bigg)_{\bar{P} \rightarrow \infty}
\nonumber\\
&=& k^2_{\perp}+(m^2+k^2_{\perp})\; \frac{(1-2x-v)^2}{4\,x\,(1-x-v)}\,. 
\label{appa4}
\end{eqnarray}
%

\subsection {Front-form case}
In the front-form case, it is appropriate to introduce quantities 
such as $\omega\cdot P= \omega^0 \,(E-\hat{n}\cdot \vec{P})$, 
$\omega\cdot p= \omega^0 \,(e-\hat{n}\cdot \vec{p}), \cdots$, 
where, up to a factor, $\omega^{\mu}$ stands for the $\xi^{\mu}$ 
introduced in the text with the condition $\omega^2=0$. 
In the ``parallel'' momentum configuration of interest here, one has
$\vec{P}_i \parallel \vec{P}_f$, while $\hat{n}$ 
is taken to be opposite to $\vec{P}_i+\vec{P}_f$. 
Defining this direction as the parallel one, one can write 
$\omega\cdot P= \omega^0 \,(E+P_{\parallel})$, 
$\omega\cdot p= \omega^0 \,(e_p+p_{\parallel}), \cdots$. 
Proceeding as above and using similar notations as 
much as possible, one can first write:
\begin{eqnarray}
&&\omega\cdot \bar{P}=\frac{1}{2}\;(\omega\cdot {P}_i+\omega\cdot 
{P}_f)\,,\hspace{1cm}
(\omega\cdot {P}_f-\omega\cdot {P}_i)=2\,y\;\omega\cdot \bar{P}\,,
\nonumber \\ &&
\omega\cdot {P}_i=(1-y)\;\omega\cdot \bar{P}\,,\hspace{2cm}
\omega\cdot {P}_f=(1+y)\;\omega\cdot \bar{P}\,. \label{appa5}
\end{eqnarray}
Making now use of relations pertinent to the ``parallel'' 
momentum configuration, different expressions of $y$ are obtained, 
allowing one to identify this quantity with the Breit frame velocity, $v$, 
already mentioned. Some intermediate steps are displayed below:  
\begin{eqnarray}
y&=&\frac{E_f+P_f-E_i-P_i}{E_f+P_f+E_i+P_i}
=\frac{ (P_f-P_i) \;\Big(1+\frac{P_f+P_i}{E_f+E_i}\Big) }{ 
(E_f+E_i)\;\Big(1+\frac{P_f+P_i}{E_f+E_i}\Big)}
=\frac{P_f-P_i}{E_f+E_i}
\nonumber \\ 
&=&\frac{(E_f+P_f-E_i-P_i)\;(E_f-P_f+E_i-P_i)}{
(E_f+P_f+E_i+P_i)\;(E_f-P_f+E_i-P_i)}
\nonumber \\ 
&=&\frac{(E_f+E_i)\;(P_f-P_i)-(E_f-E_i)\;(P_f+P_i)}{
4\,M^2+Q^2}
\nonumber \\ 
&=&\sqrt{\frac{(E_f+E_i)\;(P_f-P_i)-(E_f-E_i)\;(P_f+P_i)}{4\,M^2+Q^2} \;\;
 \frac{P_f-P_i}{E_f+E_i}}
\nonumber \\ 
&=&\sqrt{\frac{(P_f-P_i)^2-(E_f-E_i)^2}{4\,M^2+Q^2} }
=\sqrt{ \frac{Q^2}{4\,M^2+Q^2} }=v\,,\label{appa6}
\end{eqnarray}
where, in order to simplify the notation and in absence of ambiguity, 
the parallel components of $\vec{P}_i$ and $\vec{P}_f$ have been denoted 
$P_i$ and $P_f$.

Introducing now the variable $x$ defined as:
\begin{equation}
\omega \!\cdot\! p= x\;\omega \!\cdot\! \bar{P}
\;\;\Big(=\omega^0 \,(e_p-\hat{n} \!\cdot\! \vec{p})
=\omega^0 \,(e_p+p_{\parallel}) \Big)
\,, \label{appa7}
\end{equation}
and using the equality, $y=v$, one can write the following relations:
\begin{eqnarray}
&&\omega \!\cdot\! p_i=\omega \!\cdot\! P_i- \omega \!\cdot\! p
= (1\!-\!x\!-v)\;\omega \!\cdot\! \bar{P}\,,\hspace{0.6cm} 
\nonumber \\
&&\omega \!\cdot\! p_f=\omega \!\cdot\! P_f- \omega \!\cdot\! p
= (1\!-\!x\!+v)\;\omega \!\cdot\! \bar{P}\,,
\nonumber \\
&&p_{\parallel}=\frac{x^2\;(\bar{E}+\bar{P}_{\parallel})^2 
-(m^2+k^2_{\perp})}{2\,x\;(\bar{E}+\bar{P}_{\parallel})}\,,
\hspace{1.0cm} e_p=\frac{x^2\;(\bar{E}+\bar{P}_{\parallel})^2 
+(m^2+k^2_{\perp})}{2\,x\;(\bar{E}+\bar{P}_{\parallel})}\,,
\nonumber \\
&&\frac{dp_{\parallel}}{dx}=\frac{e_p}{x}\,, \hspace{5.0cm}
\frac{d\vec{p}}{e_p}=d^2k_{\perp}\;\frac{dx}{x} \, ,
\nonumber \\
&& \frac{\omega \!\cdot\! (p_f+p)\;\;\omega \!\cdot\! (p_i+p)}{
(2\,\omega \!\cdot\! p_f) \;(2\,\omega \!\cdot\! p_i) }= 
\frac{1-v^2}{4\,\Big((1-x)^2-v^2\Big)}\, ,
\nonumber \\
&&\frac{2\,(p_f+p_i) \!\cdot\! \omega  }{ 
(p_f+p_i+2\,p) \!\cdot\!  \omega}=2\,(1-x) \,.
\label{appa8}
\end{eqnarray}
The argument of the wave function in the case of the initial state 
is now given by: 
\begin{eqnarray}
k_i^2&=&-\frac{1}{4}\;(p_i-p)^2=\frac{k^2_{\perp}}{2}-\frac{m^2}{2}
+\frac{m^2+k^2_{\perp}}{4}\;
\Big(\frac{\omega \!\cdot\! p}{\omega \!\cdot\! p_i} +\frac{\omega \!\cdot\! 
p_i}{\omega \!\cdot\! p}\Big)
\nonumber \\
&=&k^2_{\perp}+\frac{m^2\!+\!k^2_{\perp}}{4}\;
\Big(\frac{\omega \!\cdot\! p}{\omega \!\cdot\! p_i} +\frac{\omega \!\cdot\! 
p_i}{\omega \!\cdot\! p}-2\Big)\nonumber \\
&=& k^2_{\perp}+(m^2\!+\!k^2_{\perp})
\;\frac{(1\!-\!2\,x\!-v)^2}{4\,x\;(1\!-\!x\!-v)}\,.
\label{appa9}
\end{eqnarray}
The r.h.s. of last equalities in Eqs. (\ref{appa8}, \ref{appa9}) are 
identical to those in Eqs. (\ref{appa2}, \ref{appa4}) but they have been 
obtained without taking the limit $\bar{P} \rightarrow \infty$. 
This in in accordance with the statement often made about 
the ``perpendicular'' momentum configuration that taking the above limit 
in an instant-form approach or working on an hyperplane with the limit 
$|\,\vec{\xi}/\xi^0\,|=|\,\hat{n}\,|=1$ (front-form) are equivalent.

\section{Making the instant-form scalar form factor, $F_0(Q^2=0)$, 
Lorentz invariant}
\label{app:b}
Considering the RQM expression of the charge form factor $F_1(Q^2)$ 
in the instant form, it is generally found that its value at $Q^2=0 $ 
is independent of the momentum of the system. With this respect, 
the particular form of the last factor in Eq. (\ref{ff1}) is essential. 
This minimal Lorentz invariance property is deeply related to current 
conservation or to a meaningful definition of the norm. It seems 
to imply that standard instant- and front-form charge form factors 
be close to each other, thus pointing to a small violation of 
Lorentz invariance, a few \% at most. In contrast, examination 
of the scalar form factor, $F_0(Q^2=0)$, generally evidences 
a sizeable one. Thus, in absence of two-body currents, discrepancies 
of a few 10\% are observed between its value at $\vec{P}=0$ and 
that other one at $|\vec{P}| \rightarrow \infty$, where it is equal 
to the standard front-form result. Determining two-body currents that 
can restore the equality, thus fulfilling a minimal restoration of Lorentz
invariance, can be done from examination of diagrams. Apart from the  fact
that this is not straightforward in a RQM framework and, in any case, 
does not apply to a phenomenological approach, the question arises of
whether there is a general method allowing one to derive currents fulfilling
the above invariance property. 

In the scalar-particle case and for an interaction model corresponding to 
the exchange of an infinite-mass boson, the underlying model allowed one 
to get the explicit expression of the correcting factor to be inserted 
in the one-body current contribution \cite{Amghar:2002jx}:
\begin{equation} 
cf_0= 1+\frac{e_p}{2\,e_i}\; \frac{(e_i+e_p)^2-E^2}{(e_i+e_p)^2}.
\label{appb1}
\end{equation}
This one has a typical off-energy shell character and, therefore, 
can be cast into the form of an interaction term, using the mass equation, 
Eq. (\ref{wf6}). In this simple model, the factor makes the instant-form form
factor $F_0(Q^2=0)$ equal to the standard front-form one. Moreover, a simple
generalization at $Q^2 \neq 0$ was found to preserve the identity
\cite{Amghar:2002jx}. With the
idea that an extension of the above correction factor could contribute to
restore Lorentz invariance, we looked at its determination. 

For our purpose, it is convenient to start from an expression of $F_0(Q^2=0)$ 
where the momenta of the  constituents are expressed in terms of the 
total momentum $\vec{P}$ and the internal variable $\vec{k}$. In the above
mentioned simple model, this quantity reads:
\begin{equation} 
F_0(Q^2=0)=\int \frac{d\vec{k}}{(2\,\pi)^3}\;
\frac{\phi^2_0(k)}{1+\vec{\tilde{k}}\cdot \vec{\tilde{P}}} \;
\Bigg(1+ \frac{1}{2} \;\frac{1-\vec{\tilde{k}}\cdot \vec{\tilde{P}}  }{
1+ \vec{\tilde{k}}\cdot \vec{\tilde{P}} } \;
\frac{4\,e_k^2-M^2}{4\,e_k^2+\vec{P}^2}\Bigg).
\label{appb2}
\end{equation}
where $\phi_0(k)\propto \sqrt{2e_k}\;(4\,e_k^2-M^2)$, 
$\tilde{k}= k/e_k$ and $\tilde{P}=P/\sqrt{4\,e_k^2+\vec{P}^2}$. Despite
appearances, it can be shown that $F_0(Q^2=0)$ does not depend 
on the momentum $\vec{P}$. A generalization of the above result to any function 
$\phi_0(k)$ supposes to modify appropriately the interaction term proportional
to $4\,e_k^2-M^2$. The simplest change consists in multiplying this term 
by a factor $g(k)$ so that $F_0(Q^2=0)$ now reads: 
\begin{equation} 
F_0(Q^2=0)=\int \frac{d\vec{k}}{(2\,\pi)^3}\;
\frac{\phi^2_0(k)}{1+\vec{\tilde{k}}\cdot \vec{\tilde{P}}} \;
\Bigg(1+  \frac{g(k)}{2}\;\frac{1-\vec{\tilde{k}}\cdot \vec{\tilde{P}}  }{  
1+\vec{\tilde{k}}\cdot \vec{\tilde{P}}} \;
\frac{4\,e_k^2-M^2}{4\,e_k^2+\vec{P}^2}\Bigg).
\label{appb3}
\end{equation}
In order to $F_0(Q^2=0)$  be independent of $\vec{P}$, it
is found that the following relation has to be fulfilled:
\begin{equation} 
\frac{g(k)}{8}\;e_k\;(4\,e_k^2-M^2)\;\phi^2_0(k)=
\int^{\infty}_k dk'\;  k'\; e_{k'} \; \phi^2_0(k'),
\label{appb4}
\end{equation}
which allows one to easily  get $g(k)$. As the above result is obtained 
for the first time and that details may be useful in other cases, we provide
here some steps. An integration over the orientation of $\vec{k}$ is first
made in Eq. (\ref{appb3}) with the result:
\begin{eqnarray}  
&&F_0(Q^2=0)=\frac{1}{2\,\pi^2}\int dk\;k^2\;\phi^2_0(k)\;
\nonumber\\ && \hspace*{0cm} \times
\Bigg(\Big(1-\frac{g(k)}{2}\;\frac{4\,e_k^2-M^2}{4\,e_k^2+\vec{P}^2}\Big) \;
\frac{1}{2\,\tilde{k}\;\tilde{P}} \; {\rm log}
\Big(\frac{1+\tilde{k}\;\tilde{P}}{1-\tilde{k}\;\tilde{P}}\Big) + 
g(k) \;\frac{e_k^2(4\,e_k^2-M^2)}{4\,e_k^4+\vec{P}^2\;m^2}  \Bigg).
\nonumber \\
\label{appb5}
\end{eqnarray}
An integration by parts has now to be done to remove the undesirable log term.
This supposes that the corresponding coefficient can be cast into the form 
of the derivative of a function as follows:
\begin{eqnarray} 
&&k^2\;\phi^2_0(k)\;
\Big(1-\frac{g(k)}{2}\;\frac{4\,e_k^2-M^2}{4\,e_k^2+\vec{P}^2}\Big) \;
\frac{1}{2\,\tilde{k}\;\tilde{P}} \nonumber \\
&&\hspace*{2cm} =-\frac{d}{dk}\Bigg(\frac{\sqrt{4\,e_k^2+\vec{P}^2}}{2\,P}
\int^{\infty}_k dk'\;  k'\; e_{k'} \; \phi^2_0(k')\Bigg).
\label{appb6}
\end{eqnarray}
By identifying the left- and right-hand sides of the equation, 
Eq. (\ref{appb4}) is obtained. After the integration by parts is performed,
the expression of $F_0(Q^2=0)$ reads:
\begin{eqnarray} 
F_0(Q^2=0)&=&\frac{1}{2\,\pi^2} 
\int dk \;\phi^2_0(k)\; \frac{g(k)}{8}\; (4\,e_k^2-M^2)\;
\nonumber \\ && \times
\Bigg(\frac{m^2(4\,e_k^2+\vec{P}^2)-4\,e_k^2\;k^2)}{4\,e_k^4+\vec{P}^2\;m^2}   
+\frac{8\,e_k^2\;k^2}{4\,e_k^4+\vec{P}^2\;m^2}\Bigg),
\label{appb7}
\end{eqnarray}
which simplifies to get the total momentum independent result:
\begin{equation} 
F_0(Q^2=0)=\frac{1}{2\,\pi^2} \int dk \; 
\phi^2_0(k)\; \frac{g(k)}{8}\; (4\,e_k^2-M^2).
\label{appb8}
\end{equation}
In the particular case of a zero-range interaction, the square 
of the wave function fulfills the relation 
$\phi^2_0(k)\propto (2e_k\;(4\,e_k^2-M^2)^2)^{-1}$, from which 
one gets $g(k)=1$, in agreement with Eq. (\ref{appb1}). In the case  
$\phi^2_0(k)\propto (2e_k\;(4\,e_k^2-M^2)^4)^{-1}$, which is not a bad
approximation for a Coulombian type interaction (Wick-Cutkosky model),
one gets  $g(k)=1/3$. It is noticed that the factor 
$(1-\vec{\tilde{k}}\cdot \vec{\tilde{P}}  )/( 
1+\vec{\tilde{k}}\cdot \vec{\tilde{P}})$ in Eq. (\ref{appb2}) 
has been conserved but one can imagine to split it into two parts 
and find a more general expression of two-body currents. 
This freedom could be used to make the instant- and front-form scalar 
form factors closer to each other at any $Q^2$, beyond the equality
at $Q^2=0$ which is achieved by the two-body currents determined above. 
The method to make the instant-form scalar form factor, $F_0(Q^2=0)$, 
independent of the momentum of the system has a rather general character. 
It could be applied for instance to the pion system whose constituents 
have a non-zero spin.

\section{Details about form factors in Dirac's point-form}
\label{app:c}
We give here a few details pertinent to the derivation of the expression 
of form factors appropriate to the implementation of a Dirac's inspired 
point form which is considered in this work, among other forms. 

\subsection {Minimal expression}
On the basis of expressions of form factors for a two-body system in other
forms, it is expected that, in the case of a single-particle current, the
expression in the point-form approach will involve integration over the
3-momenta of the struck and spectator particles, constrained by relations of
these 3-momenta to the total momentum.  A minimal expression that evidences 
Lorentz invariance reads: 
\begin{eqnarray}
F(q^2)&\propto& \int \frac{d\vec{p}}{2\,e_p\;(2\,\pi)^3} \;\;\;
\frac{d\vec{p}_i}{2\,e_i\;(2\,\pi)^3} \;\;\; 
\frac{d\vec{p}_f}{2\,e_f\;(2\,\pi)^3} \;\frac{ (2\,\pi)^6}{\pi^2}
\nonumber \\ & & \hspace*{-1cm}\times \;
\delta\Big((p_i\!+\!p\!-\!P_i)^2\Big) \;
\delta\Big((p_i\!+\!p\!-\!P_i) \cdot (p_f\!+\!p\!-\!P_f)\Big) \; 
\delta\Big((p_f\!+\!p\!-\!P_f)^2\Big)\;
\nonumber \\ & & \times \;
M\,\sqrt{e_{k_i}}\;\phi_0(k^2_i) \;\; M\,\sqrt{e_{k_f}}\;\phi_0(k^2_f) \cdots,
\label{appc1}
\end{eqnarray}
where the dots account for factors pertinent to the current describing the
interaction with an external probe. The two  functions, 
$\delta\big((p_i+p-P_i)^2\big)$ and  $\delta\big((p_f+p-P_f)^2\big)$,  
are part of the wave functions for the initial and final states
\cite{Desplanques:2004rd}. The middle one, 
$\delta\big((p_i+p-P_i) \cdot (p_f+p-P_f)\big)$, stems from the integration 
over the coordinate at the interaction point with the external probe, 
constrained to be on a hyperboloid. Involving plane waves relative 
to the struck particle and the probe one, it provides a function, 
$\delta\big((p_i-p_f+q)^2\big)$, which can be rearranged into the above one 
taking into account the other $\delta(\cdots)$ functions and the 4-momentum 
conservation relation, $P^{\mu}_i-P^{\mu}_f+q^{\mu}=0$. It is noticed 
that the introduction of a 4-vector at the r.h.s. of Eq. (\ref{appc1}), 
$p^{\mu}$ or $p^{\mu}_i+p^{\mu}_f$ for instance, would produce the 
appearance of a 4-vector at the l.h.s. with the correct transformation
properties under a Lorentz transformation, similarly to the earlier 
``point form''. The structure of the integrand is however quite different 
(compare with Eq. 8 in Ref. \cite{Desplanques:2001zw} or Eq. 42 
in Ref. \cite{Amghar:2002jx}).

\subsection {Removing the $\delta(\cdots)$ functions}
In the following, we transform the above expression into a one where
integrations over the various  $\delta(\cdots)$ functions are performed. 
Introducing vectors $\vec{u}_i$ and $\vec{u}_f$, the above expression 
can first be written:
\begin{eqnarray}
F(q^2)&\propto& \int \frac{d\vec{p}}{2\,e_p\;(2\,\pi)^3} \;\;\;
\frac{d\vec{p}_i}{2\,e_i\;(2\,\pi)^3} \;\;\; 
\frac{d\vec{p}_f}{2\,e_f\;(2\,\pi)^3} 
\nonumber \\ & & \times
\int d\vec{u}_i \; d\vec{u}_f \; \delta(u_i^2-1)\;
\delta(1-\vec{u}_i\cdot\vec{u}_f) \; \delta(u_f^2-1)\; 
\frac{(2\,\pi)^6 }{\pi^2}
\nonumber \\ & &  \times \;
\delta\Big(\vec{p}_i\!+\!\vec{p}\!-\!\vec{P}_i-\vec{u}_i\;(e_i\!+\!e_p\!-\!E_i)\Big) \; 
\delta\Big(\vec{p}_f\!+\!\vec{p}\!-\!\vec{P}_f-\vec{u}_f\;(e_f\!+\!e_p\!-\!E_f)\Big) 
\nonumber \\ & &  \times \;
M\,\sqrt{e_{k_i}}\;\phi_0(k^2_i) \;\; M\,\sqrt{e_{k_f}}\;\phi_0(k^2_f) \cdots.
\label{appc2}
\end{eqnarray}
Taking advantage of the fact that the $\delta(\cdots)$ functions involving 
the $\vec{u}$ variable can be transformed for a part into 
a $\delta(\vec{u}_i-\vec{u}_f)$ function, the above expression
can be successively written after performing various integrations: 
\begin{eqnarray}
F(q^2)&\propto& \int \frac{d\vec{p}}{2\,e_p\;(2\,\pi)^3} \;\;
\frac{d\vec{p}_i}{2\,e_i\;(2\,\pi)^3} \;\; 
\frac{d\vec{p}_f}{2\,e_f\;(2\,\pi)^3} 
\nonumber \\ & &  \times \;
\int d\vec{u}_i \;d\vec{u}_f \; \delta(u_i^2\!-\!1)\; 
 \delta(\vec{u}_i\!-\!\vec{u}_f) \; \frac{(2\,\pi)^6}{\pi}
\nonumber \\ & &  \times 
\delta\Big(\vec{p}_i\!+\!\vec{p}\!-\!\vec{P}_i-\vec{u}_i\;(e_i\!+\!e_p\!-\!E_i)\Big) \; 
\delta\Big(\vec{p}_f\!+\!\vec{p}\!-\!\vec{P}_f-\vec{u}_f\;(e_f\!+\!e_p\!-\!E_f)\Big) 
\nonumber \\ & &  \times \;
M\,\sqrt{e_{k_i}}\;\phi_0(k^2_i) \;\; M\,\sqrt{e_{k_f}}\;\phi_0(k^2_f) \cdots
\nonumber \\ 
&& \hspace*{-1.5cm}\propto \int \frac{d\vec{p}}{2\,e_p\;(2\,\pi)^3} \;\;\;
\frac{d\vec{p}_i}{2\,e_i\;(2\,\pi)^3} \;\;\; 
\frac{d\vec{p}_f}{2\,e_f\;(2\,\pi)^3} 
\;\;\; d\vec{u} \; \delta(u^2\!-\!1)\;\frac{(2\,\pi)^6}{\pi}
\nonumber \\ & & \hspace*{-1.0cm} \times  \;
\delta\Big(\vec{p}_i\!+\!\vec{p}\!-\!\vec{P}_i-\vec{u}\;(e_i\!+\!e_p\!-\!E_i)\Big) \; 
\delta\Big(\vec{p}_f\!+\!\vec{p}\!-\!\vec{P}_f-\vec{u}\;(e_f\!+\!e_p\!-\!E_f)\Big) 
\nonumber \\ & & \hspace*{-1.0cm} \times \;
M\,\sqrt{e_{k_i}}\;\phi_0(k^2_i) \;\; M\,\sqrt{e_{k_f}}\;\phi_0(k^2_f) \cdots
\nonumber \\ & & \hspace*{-1.5cm} \propto 
 \int \frac{d\vec{p}}{2\,e_p\;(2\,\pi)^3} \;\;
\frac{M}{2\,u \!\cdot\! p_i} \;\; 
\frac{M}{2\,u \!\cdot\! p_f} 
\;\;  \frac{d\vec{u}}{\pi}\; \delta(u^2\!-\!1)\;
\sqrt{e_{k_i}}\;\phi_0(k^2_i) \;\; \sqrt{e_{k_f}}\;\phi_0(k^2_f) \cdots
\nonumber \\ & &\hspace*{-1.5cm}\propto  
\int \frac{d\vec{p}}{2\,e_p\;(2\,\pi)^3} \;\;
\frac{M}{2\,u \!\cdot\! p_i} \;\; 
\frac{M}{2\,u \!\cdot\! p_f} 
\;\;  \frac{d\hat{u}}{4\,\pi}\; 
\sqrt{2\,e_{k_i}}\;\phi_0(k^2_i) \;\; \sqrt{2\,e_{k_f}}\;\phi_0(k^2_f) \cdots,
\label{appc3}
\end{eqnarray}
where $u \cdot p_i=u \cdot (P_i-p)$, $u \cdot p_f=u \cdot (P_f-p)$.
The Lorentz invariance of the last expressions (dots put apart) is not 
straightforward though the property stems from the starting point,  
Eq. (\ref{appc1}).  In order to generalize the above expression and specify 
the dots, we show directly on the last expression how Lorentz invariance 
is fulfilled despite it involves the $\vec{u}$ variable. 

\subsection {Lorentz invariance of expressions with integration over 
$\vec{u}$}
Invariance under rotations being straightforward, the Lorentz transformation 
of interest here involves boosts in some direction represented by 
a vector $\vec{V}$. Introducing the notation $V^0=(1+\vec{V}^2)^{1/2}$, 
it is defined as:
\begin{eqnarray}
&&x^0 \rightarrow x^0\;V^0- \vec{x} \cdot \vec{V},
\\ \nonumber 
&&\vec{x} \rightarrow \vec{x} + \frac{\vec{x} \cdot \vec{V}}{V_0+1}\vec{V}
-x^0\;\vec{V}.
\label{appc4}
\end{eqnarray}
Under the above transformation, a seemingly Lorentz
scalar quantity like $u \cdot X$ transforms as follows:
\begin{eqnarray}
u \cdot X=X^0-\hat{u}\cdot\vec{X} &\rightarrow& X^0\;V_0+ \vec{X}\cdot\vec{V}
-\hat{u} \cdot \Big(\vec{X} 
+\vec{V} \; (\frac{\vec{V}\cdot\vec{X}}{V_0+1}+X^0) \Big)
\nonumber \\
&&=(V_0-\hat{u}\cdot\vec{V} )\;X^0
-\Big(\hat{u} +\vec{V}\;(\frac{\hat{u}\cdot\vec{V}}{V_0+1}-1)\Big)
 \cdot\vec{X}
\nonumber \\
&&=(V_0-\hat{u}\cdot\vec{V}) \; u' \cdot X\, , 
\label{appc5}
\end{eqnarray}
where, at the last line, we introduce a new 4-vector $u'\,^{\mu}$ defined as:
\begin{equation}
u'\,^0=1,\;\;\;\vec{u}\,'= \frac{1}{V_0-\hat{u}\cdot\vec{V}}\;
\Big(\hat{u}+\vec{V}\;(\frac{\hat{u}\cdot\vec{V}}{V_0+1}-1)\Big)=\hat{u}'\, ,
\label{appc6}
\end{equation} 
The last equality can be traced back to the relation 
$\vec{u}\,'\,^2=\vec{u}\,^2=1$,
which is expected from the Lorentz-invariant condition 
$u' \cdot u'=u \cdot u=0$. 

In order to determine how the volume integration in Eq. (\ref{appc3}), 
$d\hat{u}$, changes under the Lorentz transformation, it is useful 
to invert Eq. (\ref{appc6}). One thus gets:
\begin{eqnarray}
 &&\frac{1}{V_0-\hat{u}\cdot\vec{V}}= V_0+\hat{u}' \cdot \vec{V} \, ,
\nonumber \\
&&\hat{u}= \frac{1}{V_0+\hat{u}'\cdot\vec{V}}\;
\Big(\hat{u}'+\vec{V}\;(\frac{\hat{u}'\cdot\vec{V}}{V_0+1}+1)\Big).
\label{appc8}
\end{eqnarray}
The Jacobian of the transformation, $\hat{u} \rightarrow \hat{u}'$, 
can also be calculated, taking into account the normalization 
$\hat{u}'\,^2=\hat{u}^2=1$:
\begin{equation}
\frac{d\hat{u}}{d\hat{u}'}= (\frac{1}{V_0+\hat{u}'\cdot\vec{V}})^2=
(V_0-\hat{u}\cdot\vec{V})^2.
\label{appc9}
\end{equation} 
In the expression of $F(q^2)$ given by the last line of Eq. (\ref{appc3}), 
the boost transformation leads to the appearance of a factor 
$(V_0-\hat{u}\cdot\vec{V})^2$ (one factor separately for the quantities 
$e_i-\hat{u}\cdot\vec{p}_i$ and $e_f-\hat{u}\cdot\vec{p}_f$ at 
the denominator). This factor cancels the one from the Jacobian, ensuring the 
Lorentz invariance of the expression ($\hat{u}$ is replaced by $\hat{u}'$, 
which can be renamed $\hat{u}$). From the above, it immediately follows that 
the Lorentz invariance property of the quantity $F(q^2)$ will not be affected
if the dot part in the integrand at the last line of Eq. (\ref{appc3}) 
contains factors depending on $u^{\mu}$ provided that they evidence a seemingly 
Lorentz-scalar form and are invariant under changing the scale of $u^{\mu}$. 
In such a case, the factors $(V_0-\hat{u}\cdot\vec{V})$
appearing in the Lorentz transformation, last line of Eq. (\ref{appc5}), 
cancel out. We stress that this simplification, which is essential 
to demonstrate the above Lorentz-invariance property, is possible 
because $u \cdot u=0$.

\subsection {Other factors: the dot part}
In considering the part involving dots in Eqs. (\ref{appc1}-\ref{appc3}), 
which was unspecified till here, we first look at the case, $q^{\mu}=0$, 
where the normalization condition, $F_1(0)=1$, should be recovered. 
For this quantity, the dots are replaced by:
\begin{equation}
\dots=\frac{2\, u \cdot (p_i+p_f)}{ u \cdot (p_i+p_f+2\,p)},
\label{appc10}
\end{equation}
which is suggested by the close relationship of the normalization to the
charge current density and is unchanged when the scale of $u^{\mu}$ is modified.
As explained elsewhere \cite{Amghar:2002jx}, the factor $(p_i+p_f)^{\mu}$ 
at the numerator could represent the interaction of the photon with 
the constituents while the quantity at the denominator represents 
the sum of the momenta of the constituents that has to be factored out 
in calculating the charge form factor 
($(p_i+p_f+2\,p)^{\mu}=(P_i+P_f)^{\mu}$ in absence of interaction). 

For our purpose, either equation  (\ref{appc1}-\ref{appc3}) could be used. 
Starting from Eq. (\ref{appc1}) for instance, and taking into account the
relation  $P^{\mu}_i=P^{\mu}_f$ for $q^{\mu}=0$, it is first noticed that the
product of the three $\delta(\cdots)$ function in the integrand can be
expressed as follows: 
\begin{eqnarray}
&& \delta \Big((p_i+p-P_i)^2\Big) \;
\delta \Big((p_i+p-P_i) \cdot (p_f+p-P_i)\Big) \; 
\delta \Big((p_f+p-P_i)^2\Big)\;
\nonumber \\
&&=\delta \Big((p_i+p-P_i)^2\Big) \;
\delta \Big((p_i+p-P_i) \cdot (p_f-p_i)\Big) \; 
\delta \Big((p_i-p_f)^2\Big)\;
\nonumber \\
&&=\delta \Big((p_i+p-P_i)^2\Big) \;\pi\; \frac{e_f}{(p_i+p-P_i)\cdot p_f}
 \;\delta \Big(\vec{p}_i-\vec{p}_f\Big)\,.
\label{appc11}
\end{eqnarray}
In writing the last line, we employed relations similar to Eqs. (74) and (75) 
of Ref. \cite{Desplanques:2004rd}, taking into account that $p_i^2=p_f^2=m^2$. 
The last  $\delta(\cdots)$ function in the above equation allows one to perform
the integration over $\vec{p}_f$ in Eq. (\ref{appc1}). The form factor 
$F_1(0)$ now reads:
\begin{eqnarray}
F_1(0) &=& \int\frac{1}{(2\,\pi)^6} \;\frac{d\vec{p}}{2\,e_p} \;\;
\frac{d\vec{p}_i}{2\,e_i} \;\;\; 
\frac{4\,\pi^2}{(p_i+p-P_i)\cdot (p_i+p)} 
\nonumber \\ & & \times \;
\delta\Big((p_i+p-P_i)^2\Big) \;
\Big(M\,\sqrt{e_{k_i}}\;\phi_0(k^2_i) \Big)^2\;\; 
\nonumber \\ & =& \int\frac{1}{(2\,\pi)^6} \;\frac{d\vec{p}}{2\,e_p} \;\;
\frac{d\vec{p}_i}{2\,e_i} \;\; 
\frac{e_{k_i}\;\phi_0^2(k^2_i)}{4\,e^2_{k_i}-M^2} \;
8\,\pi^2 \delta\Big((p_i+p-P_i)^2\Big) \;M^2\, ,
\label{appc12}
\end{eqnarray}
which, apart from notations, can be seen to be identical to Eq. (31) given 
in Ref. \cite{Desplanques:2004rd}. After making a change of variable described
in this last work, one also finds:
\begin{equation}
F_1(0)=\int \frac{d\vec{k}}{(2\pi)^3}\;\phi^2_0(k)  
\int \frac{d\vec{u}}{2\,\pi }\; \delta(1-\vec{u}\,^2)\;
\frac{M^2}{(u \!\cdot\! P_i)^2}=1\, .
\label{appc13}
\end{equation}
It is noticed that Eq. (\ref{appc12}) can be easily recovered from 
the second relation of Eq. (\ref{appc3}). For $P^{\mu}_i=P^{\mu}_f$, the second
$\delta(\cdots)$ function can be readily transformed into a  
$\delta \big(\vec{p}_i-\vec{p}_f\big)$ function, which allows one to
make the integration over $\vec{p}_f$. Accounting for the appropriate 
factors, the integration over $\vec{u}$ is easily performed using 
the other 3-dimensional $\delta(\cdots)$ function. 

At first sight, expressions (\ref{appc1}-\ref{appc3}) together with the
appropriate choice for the dots, given by Eq. (\ref{appc10}) for the charge
form factor, by 1 for the Lorentz-scalar one, could be used for 
a non-zero momentum transfer since the correct charge form factor 
at $q^2=0$ is obtained.  However, one can imagine to insert 
extra factors such as $u \cdot P_i / u \cdot P_f $ 
or $u \cdot P_f / u \cdot P_i $  
in the dots part of Eqs. (\ref{appc1}-\ref{appc3}) since these ones 
preserve Lorentz invariance and reduce to 1 at zero momentum transfer, 
allowing one to fulfill the above limit. While trying to fix this 
extra factor, we have in mind that the structure for the currents should 
be close to each other in different forms so that to avoid some bias
in comparing their predictions. From considering Eq. (\ref{ff1}), a minimal
factor, corresponding to the quantity 
$\xi_f \cdot (p_f+p)\;\;\xi_i \cdot (p_i+p)$ appearing at its numerator,  
is given by $u \cdot P_i \;u \cdot P_f$. The full factor should be $u^{\mu}$-scale
independent and, therefore, this quantity has to be divided by a factor 
that is bilinear in $u^{\mu}$. As the expression of the norm given in Eq.
(\ref{appc13}) suggests, it could be either  $(u \cdot P_i)^2$ 
or $(u \cdot P_f)^2$. These possibilities correspond to the expectation 
that, in the c.m., the integration over $\hat{u}$ should be made 
isotropically. Moreover, what is isotropic for the initial state may 
not be for the final state and vice versa. The ambiguity has probably 
its origin in a partial treatment of the ``time'' evolution of the 
interaction with an external probe. It is illustrated here by the consideration 
of the following $\delta(\cdots)$ function that could appear 
in a more complete treatment. Depending on how the limit of the energy
conservation is taken, one could get different results:
\begin{eqnarray}
\delta(\vec{P}_i-\vec{P}_f+\vec{q}-\hat{u} \;(E_i-E_f+q^0))
&=&\delta(\vec{P}_i-\vec{P}_f+\vec{q}) 
\nonumber \\ 
&=&\frac{E_i}{u \cdot P_i}\; \delta(\vec{P}_i-\vec{P}_f+\vec{q})
\nonumber \\ 
&=&\frac{E_f}{u \cdot P_f}\; \delta(\vec{P}_i-\vec{P}_f+\vec{q}) \, .
\label{appc14}
\end{eqnarray}
Of course, physical results should not depend on either expression. 
Other examples can be encountered, in relation with off-energy shell effects. 
In the case of the electromagnetic interaction for instance, 
one can consider that the photon is emitted either from  the initial 
or final state. Accordingly, one could have relations like 
$P_f^2=(P_i+q)^2$ and $P_i^2=(P_f-q)^2$. In principle, the corresponding
contributions should be the same. However, in an incomplete calculation, 
they may differ.   In analogy with this example, the two contributions 
should be considered on an equal footing.  The dot part of 
Eqs. (\ref{appc1}-\ref{appc3}) can therefore be replaced as follows:  
\begin{eqnarray}
&& F_1(Q^2) \rightarrow \;\; \dots= 
\frac{u \cdot P_i \; u \cdot P_f}{2}\;
\Big( \frac{1}{(u \cdot P_f)^2} +\frac{1}{(u \cdot P_i)^2} \Big)
\; \frac{2\, u \cdot (p_i+p_f)}{ u \cdot (p_i+p_f+2\,p)}\;,
\nonumber \\ 
&& F_0(Q^2) \rightarrow \;\; \dots= \frac{u \cdot P_i \; u \cdot P_f}{2}\;
\Big( \frac{1}{(u \cdot P_f)^2} +\frac{1}{(u \cdot P_i)^2} \Big)\, .
\label{appc15}
\end{eqnarray}
Actually, form factors calculated here turn out 
to be independent of which term is consi\-de\-red in the above expression. 
We can therefore omit one of them together with the front factor $1/2$, 
what is made  in the text, Eq. (\ref{ff7}). It is noticed that this expression 
allows one to recover the expected asymptotic behavior of the form factor 
in the Born-amplitude approximation \cite{Desplanques:2003nk,Desplanques:2004rd}.
In our opinion, this result, which could be used, the other way round, 
to discriminate among choices for the factor in place of 
$(u \cdot P_f)^{-2} + (u \cdot P_i)^{-2}$ in Eq. (\ref{appc15}), 
is not fortuitous. The asymptotic Born amplitude is the sum of two terms 
where one of the initial or final states is on mass-shell while the other one 
is not. For an off-mass-shell initial state for instance and an on-mass-shell 
final one, the absence of interaction in this last state discards dependence 
on $u \cdot P_f$, leaving only ($u \cdot P_i)^{-2}$ as a possible choice. 
The symmetry between the initial and final states then suggests to take 
the combination $(u \cdot P_f)^{-2} + (u \cdot P_i)^{-2}$.

%

\end{document}